\documentclass[11pt,a4paper]{article}
\pdfoutput=1
\usepackage{jheppub}
\usepackage{gensymb}
\usepackage{subfigure}
\usepackage{amssymb,amsmath}
\usepackage{graphicx}
\usepackage{color}
\usepackage{cancel}
\usepackage[colorlinks=true
,urlcolor=blue
,citecolor=blue
,linkcolor=blue
,pagecolor=blue
,linktocpage=true
,pdfproducer=medialab
]{hyperref}
 \usepackage[numbers]{natbib}
\usepackage{notoccite}
\makeatletter \renewcommand{\@dotsep}{10000} \makeatother
\def\be{\begin{equation}}
\def\ee{\end{equation}}
\def\bea{\begin{eqnarray}}
\def\eea{\end{eqnarray}}
\def\bi{\begin{itemize}}
\def\ei{\end{itemize}}

\def\to{\rightarrow}

\usepackage[nodisplayskipstretch]{setspace}

\def\mgut{M_{\rm GUT}}

\newcommand{\beq}{\begin{equation}}
\newcommand{\eeq}{\end{equation}}

\begin{document}

\begin{titlepage}
\pagestyle{empty}

\begin{flushright}
OSU-HEP-20/16
\end{flushright}

\vspace*{0.2in}
\begin{center}
{\Large \bf    Proton Lifetime in Minimal SUSY \boldmath{$SU(5)$}\\[0.1in] in Light of LHC Results
  }\\
\vspace{1cm}
{\bf  K.S. Babu$^{a,}$\footnote{E-mail: babu@okstate.edu},
Ilia Gogoladze$^{b,}$\footnote{E-mail: ilia@physics.udel.edu;} {\rm and}
Cem Salih $\ddot{\rm U}$n$^{c,d,}\hspace{0.05cm}$\footnote{E-mail: cemsalihun@uludag.edu.tr}}
\vspace{0.5cm}

{\it
$^a$Department of Physics, Oklahoma State University, Stillwater, OK, 74078, USA \\
$^b$Department of Physics and Astronomy,
University of Delaware, Newark, DE 19716, USA \\
$^c$Department of Physics, Bursa Uluda\~{g} University, TR16059 Bursa, Turkey \\
$^d$ Departamento de Ciencias Integradas y Centro de Estudios Avanzados en F\'{i}sica Matem\'aticas y Computación, Campus del Carmen, Universidad de Huelva, Huelva 21071, Spain}

\end{center}

\vspace{0.5cm}
\begin{abstract}

 We examine proton decay mediated by color-triplet Higgsinos in minimal supersymmetric $SU(5)$ grand unified theory in light of the discovery of the Higgs boson and the absence of SUSY signals at the LHC. We pay special attention to various threshold effects arising from Planck-suppressed operators that affect the color-triplet Higgsino mass and also { allow for correcting} the wrong mass relations for the light fermions.  Our analysis allows for a non-universal SUSY spectrum with the third family sfermions having a separate mass compared to the first two families.  We identify the allowed parameter space of the model and show that the SUSY scalar masses are constrained by current limits from proton lifetime to be above 5 TeV, while the glunio, Wino and the Higgsinos may be within reach of the LHC. When the SUSY scalar masses are required to be $\leq 30$ TeV, so that they are within reach of next generation collider experiments, we find that proton lifetime for the decay $p \rightarrow \overline{\nu} K^+$ is bounded by $\tau(p \rightarrow \overline{\nu} K^+) \leq 1.1 \times 10^{35}$ yrs.

\end{abstract}
\end{titlepage}

\tableofcontents

\noindent \hrulefill

\section{Introduction}
\label{sec:intro}

Supersymmetric (SUSY) grand unified theories (GUTs) based on the gauge group $SU(5)$ are attractive extensions of the Standard Model \cite{Dimopoulos:1981zb,Sakai:1981gr}.  They are supported by the meeting of the three gauge couplings  at an energy scale of $M_{\rm G} = 2 \times 10^{16}$ GeV, assuming that the SUSY particles have TeV scale masses.  These theories also provide an excellent dark matter candidate in the neutralino LSP (lightest SUSY particle).  SUSY is a necessary ingredient of string theory, which is the best candidate we have for a theory of quantum gravity.  SUSY can provide a solution to the gauge hierarchy problem, if the SUSY particles have masses not much above the TeV scale -- a feature under siege by the non-observation of SUSY particles at the Large Hadron Collider (LHC).  GUTs \cite{Pati:1974yy,Georgi:1974sy}, with or without SUSY are attractive on several other grounds: they explain the observed quantization of electric charge, they organize quarks and leptons into common and simple multiplets, and they provide an understanding of the anomaly cancellation. The hallmark prediction of this class of theories is that the proton should ultimately decay, with a lifetime predicted to be not far from the current limits and perhaps within reach of ongoing and forthcoming experiments.

In SUSY GUTs, the dominant contribution to proton decay amplitude arises from color-triplet Higgsinos which are the GUT partners of the Higgs boson \cite{Weinberg:1981wj,Sakai:1981pk,Langacker:1980js,Nath:2006ut}. The decay rate from these $d=5$ operators scales as $(M_{H_C})^{-2}$, where $M_{H_C}$ denotes the mass of the color-triplet Higgsino,  which is typically more dominant over the $d=6$ gauge  boson mediated proton decay rate, which scales as $(M_{V})^{-4}$ with $M_V$ being the GUT scale mass of the gauge bosons.  In spite of suppressions from light fermion family Yukawa couplings, and a loop factor that is needed for {dressing} of the effective $d=5$ operators, these dominant Higgsino mediated operators have been in some tension with experimental limits on proton lifetime, $\tau(p \rightarrow \overline{\nu} K^+) \geq 5.9 \times 10^{33}$ yrs. \cite{Abe:2014mwa}, for typical  parameters in any SUSY GUT.  This raises the question as to the viability of minimal SUSY $SU(5)$, especially in view of the discovery of the Higgs boson \cite{Aad:2012tfa,Chatrchyan:2012ufa} as well as improved limits from SUSY particle searches from the Large Hadron Collider (LHC) experiments \cite{Aaboud:2017dmy,Sirunyan:2017cwe}.  These results from the LHC do provide important restrictions on the lifetime of the proton within SUSY $SU(5)$. The purpose of this paper is to undertake a careful quantitative analysis to address this question.

We define minimal SUSY $SU(5)$ as a theory with the minimal particle content and a renormalizable superpotential, but one which allows the inclusion of Planck-suppressed non-renormalizable operators.  These non-renormalizable operators, which appear with a suppression factor $(M_{\rm G}/M_{\rm Pl}) \sim 10^{-2}$, will only play a sub-leading role in symmetry breaking and fermion mass generation.  Sometimes minimal SUSY $SU(5)$ is defined without the presence of the non-renormalizable operators;  however, in this case, the masses of light fermions predicted by the theory are inconsistent with experimental observations. New ingredients would be needed to correct the wrong mass relations, thereby invalidating the minimal theory.  The presence of Planck-suppressed operators, which are presumably present in any theory, would correct the wrong fermion mass relations without the need to introduce new particles. While we allow for various types of Planck-suppressed operators, we take them to be small, which is the case when the GUT scale, $M_{\rm G} = 2 \times 10^{16}$ GeV, is clearly separated from the (reduced) Planck scale, $M_{\rm Pl} = 2.4 \times 10^{18}$ GeV by two orders of magnitude.

If all the SUSY particles have masses below about 5 TeV  -- a mass range that is being probed currently by the LHC experiments -- and if the GUT scale threshold effects are negligible, there is no room for minimal SUSY $SU(5)$ to be consistent with proton lifetime limits. However, the assumptions made to lead to such a conclusion are suspect.  First, as already noted, the minimal SUSY $SU(5)$ theory in its renormalizable version leads to wrong relations to the fermion masses, especially for those in the first two families.  Since the Higgsino mediated $d=5$ proton decay rate is intimately tied to the Yukawa couplings of the light fermions, any new effect that corrects the wrong mass relations would also modify proton lifetime estimates.  Second, the assumption that SUSY particles have masses not exceeding about 5 TeV may not be justified; it originates from naturalness arguments (and the desire to observe the particles at the LHC), which may be flawed. It would therefore be interesting to evaluate proton lifetime constraints on the model parameters allowing for the presence of Planck induced effects, and with SUSY particles heavier than 5 TeV, which is what we undertake in this paper. If the SUSY scalar masses are limited to be less than about 30 TeV, they could be within reach of next generation collider experiments, and they may still offer partial solution to the gauge hierarchy problem.  We adopt this range of scalar masses in our analysis. When LHC constraints are folded in  and the current proton lifetime limit is imposed, we find that at least some of the SUSY scalars must have masses above 14 TeV, {while} the gluino, the Wino and the Higgsinos are within reach of the LHC. We also find that the lifetime of the proton for the decay $p \rightarrow \overline{\nu} K^+$ is bounded from above to be $\tau(p \rightarrow \overline{\nu} K^+) \leq 1.1 \times 10^{35}$ yrs.

In our analysis we stick with the particle content of the minimal SUSY $SU(5)$ model \cite{Dimopoulos:1981zb,Sakai:1981gr,Chamseddine:1982jx}. This includes three chiral superfields belonging to ${\bf 10} + {\bf \overline{5}}$ and Higgs superfields belonging to ${\bf 24} + {\bf 5} + {\bf \overline{5}}$ of $SU(5)$. Furthermore, we assume $R$-parity conservation.  The novel features of our analysis are as follows:

{\bf 1.}  We correct the wrong mass predictions of minimal SUSY $SU(5)$ arising from the asymptotic relation $M_d^0 = (M_\ell^0)^T$ connecting the down quark and charged lepton mass matrices at the GUT scale.  This relation would imply, in particular, that $m_\mu^0/m_s^0=1$ for the muon to strange quark mass ratio at the GUT scale, which is off by a factor of 4.4 compared to its experimental value.  If this wrong relation is used for proton decay calculation, the lifetime would be over-estimated by a factor of $(4.4)^2 = 19.4$.  The inclusion of the correct masses therefore further constrains the allowed parameter space of the theory. We accommodate the correct masses by including Planck-suppressed operators { of specific flavor structure} that can potentially arise from quantum gravity  \cite{Ellis:1979fg}. Although suppressed by a factor $(M_{\rm G}/M_{\rm Pl}) \simeq 10^{-2}$, these operators are adequate  to correct the wrong relations among light fermion masses.

{\bf 2.}  We allow for Planck-suppressed operators in the symmetry breaking sector of the superpotential as sub-leading corrections, which modify the value of the color-triplet Higgsino mass, a crucial ingredient for proton lifetime estimate. We also allow for Planck-suppressed operators in the gauge kinetic term, which modifies the interconnections between various GUT scale particle masses.  The dimensionless  coefficients of these Planck-suppressed operators will be taken to be of order unity.

{\bf 3.} We allow for the third  family squarks and sleptons to have a separate mass at the GUT scale compared to the first two families.  { Phenomenology of such a SUSY breking scenario  has been studied under the name NUHM2 \cite{Ellis:2002wv}}. This is { also} justified by flavor symmetry arguments compatible with GUTs as illustrated in the context of symmetry-based MSSM (sMSSM) \cite{Babu:2014sga,Babu:2014lwa}.  This relaxes proton decay constraints somewhat. It should be noted that with three family universality assumption, the LHC limits on SUSY scalar masses are somewhat more constraining compared to the $2+1$ splitting of masses adopted here.

There have been various approaches to address the $d=5$ proton decay issue within SUSY $SU(5)$.  Ref. \cite{Murayama:2001ur} takes the renormalizable theory at face value and argues that not even raising the SUSY scalar masses of the first two families could salvage the minimal SUSY $SU(5)$ model.  While this is true, the assumption of not allowing Planck-suppressed operators in the superpotential adopted in Ref. \cite{Murayama:2001ur}  appears to be too rigid to us. That assumption also leaves the wrong relation $m_\mu^0/m_s^0 = 1$ uncorrected, which would make the theory not fully consistent. Indeed, our analysis shows that consistent parameter space exists with the inclusion of Planck-suppressed operators with relatively small magnitudes. The authors of Ref. \cite{Bajc:2002bv}, on the other hand, admit arbitrary and large threshold corrections in the superpotential, which would allow for the GUT scale to be raised to values well above $M_{\rm G} = 2 \times 10^{16}$ GeV, even as large as $M_{\rm Pl}$.  This can be realized if the remnants of $SU(5)$ symmetry breaking have intermediate scale masses, which may occur when the Planck-suppressed operators dictate the GUT symmetry breaking \cite{Bachas:1995yt,Chkareuli:1998wi,Emmanuel-Costa:2003szk}.
In this case Planck-suppressed corrections even of higher order become important, making the theory not predictive as regards gauge coupling unification and proton lifetime. Our approach here is somewhere in between; we do rely on Planck-suppressed operators, but they remain small compared to the renormalizable operators. The observed unification of gauge couplings within the MSSM is only modified slightly in this case. There are of course other ways of correcting the fermions mass relations, such as introducing Higgs multiplets in the ${\bf 45}+\overline{\bf 45}$ of $SU(5)$ \cite{Georgi:1979df} -- potentially with large GUT scale threshold effects in the SUSY context, or by introducing a vector-like fermion in the ${\bf 5}+\overline{\bf 5}$ of $SU(5)$ with smaller threshold effects \cite{Babu:2012pb}. Our analysis differs from these variants in that we stay with the spectrum of minimal SUSY $SU(5)$. Ref. \cite{Bajc:2015ita} has studied minimal SUSY $SU(5)$ theory in its renormalizable version, allowing for the fermion masses to be corrected by SUSY threshold effects.  Here it has been shown that if the masses of the SUSY particles are of order $(10^2 - 10^4)$ TeV, the model can be made realistic.  In contrast to this work, we stay with SUSY sclalar masses to be at most 30 TeV. Recently proton lifetime and SUSY spectrum has been analyzed  including constraints from the LHC and from SuperKamiokande in Ref. \cite{Ellis:2016tjc,Ellis:2019fwf}. Our analysis is similar in spirit, but we differ by the inclusion of items ${\bf 1}$ and {\bf 3} listed above, as well as the restriction of scalar masses $\leq 30$ TeV that we have adopted.

The rest of the paper is organized as follows. In Sec. \ref{sec2} we present the minimal SUSY $SU(5)$ setup in the renormalizable version. In Sec. \ref{sec3} we include Planck-suppressed  threshold effects of various types and identify the allowed mass scale of the color-triplet Higgsino.  In Sec. \ref{sec:scan} we summarize our scanning procedure and outline the various experimental constraints used.  In Sec. \ref{sec:funparam}, we present our results, including constraints from proton lifetime.  Sec. \ref{sec:d=6} has an update on the sub-leading $d=6$ gauge boson mediate proton decay.  In Sec. \ref{sec:conc} we conclude. Details of the $d=5$ proton decay calculations adopted are presented in the Appendix.

\section{Minimal SUSY SU(5): The General Setup}\label{sec2}

In this section we summarize the framework of minimal SUSY $SU(5)$ in its renormalizable version. We derive ranges for the masses of GUT scale particles consistent with low energy measurements on the Higgs boson mass, radiative electroweak symmetry breaking requirement with a neutral LSP, $B$ meson decay constraints, and lower limits on SUSY particle masses from the LHC.  We adopt a universal mass for the first two family squarks and sleptons at the GUT scale $m_{0_{1,2}}$,  and a separate mass for the third family  $m_{0_{3}}$ in our analysis. Such a spectrum is motivated in general supergravity theories with a flavor symmetry that treats the first two families as a doublet of a non-Abelian flavor group, referred to as symmetry-based MSSM (sMSSM) \cite{Babu:2014sga,Babu:2014lwa}.  Each of these mass parameters  ($m_{0_{1,2}}$ and $m_{0_{3}}$)  is allowed to take values as large as 30 TeV. This imposed upper limit is motivated by a partial solution to the hierarchy problem, as well as the potential to discover these particles at the next generation colliders. Gaugino mass unification is assumed, as is required in a GUT, with $M_{1/2} \leq 2$ TeV imposed, corresponding to a gluino mass of  6 TeV.  A SUSY spectrum with $M_{1/2} \ll m_{0_{1,2,3}}$ is preferred from $d=5$ proton decay constraints, which justifies the relatively low value of $M_{1/2}$ used. This range of $M_{1/2}$ also can provide a WIMP dark matter in the form of a neutralino.  Such a spectrum also leaves the possibility open for the gauginos to be discovered at the high luminosity run of the LHC.  The full range of MSSM parameters used are shown in Eq. (\ref{paramSP}) of Sec. \ref{sec:scan}.
The results derived in this section will be improved in the next section where we undertake a similar analysis, but including various Planck-suppressed non-renormalizable operators as sub-leading corrections to the theory.

\subsection{The renormalizable SUSY $SU(5)$}

Fermions of each family are assigned to ${\bf 10} + {\bf \overline{5}}$ representations of $SU(5)$. We denote these fields as $\Psi_i^{ab} = -\Psi_i^{ba}$ and $\Phi_{ia}$ respectively, where $(a,b)$ are $SU(5)$ indices, while $i$ is the family index.   These fields can be expressed in matrix form as:
\begin{eqnarray}
\Psi =\frac{1}{\sqrt{2}} \left(\begin{matrix} 0 & u_3^c & -u_2^c & u_1 & d_1 \\
-u_3^c & 0 & u_1^c & u_2 & d_2 \\
u_2^c & -u_1^c & 0 & u_3 & d_3 \\
-u_1 & -u_2 & -u_3 & 0 & e^c \\
-d_1 & -d_2 & -d_3 & -e^c & 0   \end{matrix}  \right),~~~\Phi =
\left(\begin{matrix} d_1^c \\ d_2^2 \\ d_3^c \\ e \\ -\nu  \end{matrix}  \right)~.
\end{eqnarray}
Here indices $1,2,3$ are the color indices, and the family index $i$ is suppressed.

The Higgs sector of minimal SUSY $SU(5)$ consists of an adjoint ${\bf 24}$ (denoted as $\Sigma$) and a ${\bf 5} + {\bf \overline{5}}$ pair (denoted as $H+ \overline{H}$).
The renormalizable superpotential of the theory involving only the Higgs fields is given by
\begin{eqnarray}
W_5=\frac{m_{\Sigma}}{2}{\rm Tr}(\Sigma^2) +\frac{1}{3}f\, {\rm Tr}(\Sigma^3)+m_H \overline{H}H +\lambda\,\overline{H} \Sigma H~.
\label{gut poten}
\end{eqnarray}
The $\Sigma$ field breaks $SU(5)$ in the SUSY limit down to the MSSM once it acquires a vacuum expectation value {(VEV)} along the SM singlet direction:
\begin{equation}
\left\langle \Sigma \right \rangle = {\rm diag.}(2,\,2,\,2,\,-3,\,-3) \times \sigma~.
\end{equation}
This also generates masses for the $X$ and $Y$ gauge bosons of $SU(5)$, having $SU(3)_c \times SU(2)_L \times U(1)_Y$ quantum numbers  $X(3,2,-5/6)$ and $Y(\overline{3},2,5/6)$, given by
\begin{eqnarray}
 M_X=M_Y=5\sqrt2 g_5 \, \sigma~.
\label{XY-boson}
\end{eqnarray}
Here $g_5$ is the unified $SU(5)$ gauge coupling, which has a numerical value of $g_5 \simeq 0.72$.

The $(3,2,-5/6)+(\overline{3},2,5/6)$ components of the {\bf 24}-Higgs multiplet are eaten up by the $X$ and $Y$ gauge bosons via the super-Higgs mechanism, leaving behind {three  physical Higgs states, a color octet $\Sigma_8(8,1,0)$, an $SU(2)_L$ triplet $\Sigma_3(1,3,0)$  and singlet $\Sigma_1(1,1,0)$ which have the following masses}
\begin{eqnarray}
M_\Sigma =M_{\Sigma_8}=M_{\Sigma_3}=\frac{5}{2}f\, \sigma=\left(\frac{1}{2\sqrt2}  \right) \left(\frac{f}{g_5}  \right) M_X~~ \nonumber \\ M_{\Sigma_1}=\frac{1}{2}f\sigma=\left(\frac{1}{10\sqrt2}  \right) \left(\frac{f}{g_5}  \right) M_X.
\label{octet-triplet}
\end{eqnarray}

The $H+\overline{H}$ fields contain the MSSM Higgs doublets $H_u$ and $H_d$, as well as color-triplet partner fields $H_C$ and $\overline{H}_C$.  It is these color-triplet fields that mediate proton decay via baryon number violating $d=5$ effective superpotential couplings.  These fields are parametrized as:
\begin{equation}
H = \left(H_{1C},\,H_{2C}\, H_{3C},\, H_u^+,\, H_u^0 \right)^T,~~~~~\overline{H} = \left(\overline{H}_{1C},\, {H}_{2C},\, \overline{H}_{3C},\, H_d^-,\,-H_d^0\right)^T~.
\end{equation}
The last two components of $H$ and $\overline{H}$ form  doublets of $SU(2)_L$, which are identified as $H_u$ and $H_d$ of MSSM respectively.
The masses of the color-triplet Higgs fields $M_{H_C}$ and the MSSM parameter $\mu$ can be read off from Eq. (\ref{gut poten}):
\begin{equation}
    M_{H_C} = m_H + 2\, \lambda\, \sigma,~~~\mu = m_H - 3\,\lambda\, \sigma~.
\end{equation}
While the $\mu$-parameter should be of order TeV for consistent phenomenology, $M_{H_C}$ should be of order the GUT scale since the color-triplet Higgsino mediates $d=5$ proton decay.  This is achieved by fine-tuning the two terms in the expression for $\mu$ to the desired value. With this fine-tuning,  $M_{H_C}$ becomes
\begin{equation}
M_{H_C} = 5\,\lambda \sigma =
\left(\frac{1}{\sqrt2}  \right) \left(\frac{\lambda}{g_5}  \right) M_X~.
\label{color-higgs}
\end{equation}

The Yukawa superpotential of the model consists {of the following terms at the renormalizable level}:
\begin{eqnarray}
W_{\rm Yuk} = \frac{1}{4} h^{ij} \epsilon_{abcde} \Psi_i^{ab}\Psi_j^{cd} H^e -\sqrt{2} f^{ij} \Psi_i^{ab} \Phi_{ja}\overline{H}_b~.
\end{eqnarray}
This can be decomposed in terms of the SM fields and the color-triplet Higgsino fields $(H_C,\,\overline{H}_C)$ as
\begin{eqnarray}
W_{\rm Yuk} &=& h^{ij} Q_i^{\alpha p} u^c_{j \alpha} H_u^q \epsilon_{pq} - f^{ij} Q_i^{\alpha p} d^c_{j \alpha} H_d^q \epsilon_{pq} - f^{ij} e_i^c L_j^p H_d^q \epsilon_{pq} \nonumber \\
&-& \frac{1}{2} h^{ij} Q_i^{\alpha p} Q_j^{\beta q} H_C^\gamma \epsilon_{pq} \epsilon_{\alpha \beta \gamma} + f^{ij} Q_i^{\alpha p} L_j^q \overline{H}_{C\alpha} \epsilon_{pq} \nonumber \\
&+& h^{ij} u^c_{i\alpha} e^c_j H_C^\alpha - f^{ij}u^c_{i \alpha} d^c_{j \beta} \overline{H}_{C\gamma} \epsilon^{\alpha \beta \gamma}~.
\label{Yukexp}
\end{eqnarray}
Here $(p,\,q)$ are $SU(2)_L$ indices, $(\alpha,\,\beta,\,\gamma)$ are $SU(3)_C$ indices, and $(i,\,j)$ are family indices.
In the standard notation of MSSM we define
\begin{equation}
    \left\langle H_u^0 \right \rangle = v_u,~~ \left\langle H_d^0 \right \rangle = v_d,~~~\tan\beta = \frac{v_u}{v_d}~.
\end{equation}
The mass matrices for up-quarks, down-quarks and charged leptons that follow from Eq. (\ref{Yukexp}) are then
\begin{equation}
M_u = h\, v_u,~~~M_d = f\, v_d,~~~M_\ell = f^T\, v_d~.
\label{mass}
\end{equation}
The last two relations of Eq. (\ref{mass}) will lead to the equality of mass eigenvalues of the down-type quarks and charged leptons at the GUT scale: $m_b^0 = m_\tau^0,\,m_s^0 = m_\mu^0,\,\mathbf{m_d^0 = m_e^0}$.  The first of these relations is approximately found to be valid when the low energy masses of $b$ quark and $\tau$ lepton are extrapolated to the GUT scale, but the last two relations are violated by large amounts.  Extrapolating the low energy values of the strange quark and muon masses to the GUT scale, their mass ratio is found to be $m_\mu^0/m_s^0 \simeq 4.4$, in conflict with the prediction that this ratio is one in the minimal $SU(5)$ theory with renormalizable Yukawa couplings of Eq. (\ref{Yukexp}). {Since the Higgsino-mediated proton lifetime critically depends on the masses of these light fermions, these wrong mass relations should be fixed in order to reliably estimate the $d=5$ proton decay rate.} In the next section we show how this can be achieved by staying within the minimal model, but allowing for Planck-suppressed operators that correct the predictions of Eq. (\ref{mass}).

By evolving the three gauge couplings $g_i$ of the Standard Model from the $Z$-boson mass scale to the GUT scale where they should unify, one can arrive at two relations among the gauge couplings at the $Z$-boson mass scale involving an effective GUT mass scale {$M_{\rm G}\equiv(M_X^2 M_\Sigma)^{1/3}$,} the color-triplet Higgsino mass $M_{H_C}$, and an effective mass scale for the supersymmmetric particles $m_{SUSY}$:
\begin{eqnarray}
(- 2 \alpha_3^{-1} - 3 \alpha_2^{-1} + 5 \alpha_1^{-1} ) (m_Z)
	&=& \frac{1}{2\pi} \left\{
		12 \, \ln \frac{M_X^2 M_\Sigma}{m_Z^3}
		+ 8 \ln \frac{m_{SUSY}}{m_Z} \right\},
		\label{M_XM_Sigma-1}\label{eq:2.13} \\
(- 2 \alpha_3^{-1} + 3 \alpha_2^{-1} - \alpha_1^{-1} ) (m_Z)
	&=& \frac{1}{2\pi} \left\{
		\frac{12}{5} \, \ln \frac{M_{H_C}}{m_Z}
		-2 \ln \frac{m_{SUSY}}{m_Z} \right\}.
		\label{M_XM_Sigma-2}
\end{eqnarray}
These relations are obtained by solving the one-loop renormalization group equations (RGE) assuming a TeV scale SUSY spectrum given by $16 \pi^2(dg_i/dt) =b_i g_i^3$ with $(b_1,\, b_2,\, b_3) = (33/5,\, 1,\, -3)$, where $t={\rm ln}\,\mu$. While Eqs. (\ref{M_XM_Sigma-1})-(\ref{M_XM_Sigma-2}) are written down with the assumption of a common SUSY particle mass, this can be easily improved with the following replacements that account for spread in the low energy spectrum \cite{Hisano:1992mh,Hisano:1992jj,Yamada:1992kv,Chkareuli:1998wi,Hisano:1994hb}:
\begin{eqnarray}
{\rm ln} \frac{M_{SUSY}}{M_Z}\rightarrow \frac{1}{2}{\rm ln}\frac{m_{\tilde g}}{m_Z}
+\frac{1}{2}{\rm ln}\frac{m_{\tilde w}}{m_Z}+
\frac{1}{4}{\rm ln}\frac{m^2_{\tilde Q_i}}{m_{\tilde{e^c_i}} m_{\tilde{u^c_i}}} +
\frac{1}{8}{\rm ln}\frac{m^2_{\tilde Q_3}}{m_{\tilde e^c_3}m_{\tilde u^c_3}}
\label{sisy-2}
\end{eqnarray}
in Eq. (\ref{M_XM_Sigma-1}) and
\begin{eqnarray}
{\rm ln} \frac{M_{SUSY}}{M_Z}\rightarrow -2{\rm ln}\frac{m_{\tilde g}}{m_{\tilde w}}
+\frac{4}{5}{\rm ln}\frac{m_{\tilde h}}{m_Z}+
\frac{1}{5}{\rm ln} \frac{m_{H}}{m_Z} -
\frac{1}{5}{\rm ln}\frac{m^3_{\tilde u_i^c} m^2_{\tilde d_i^c} m_{\tilde e_i^c}}{m^4_{\tilde Q_i}m^2_{\tilde L_i}}
-\frac{1}{10} {\rm ln}\frac{m^3_{\tilde u_3^c} m^2_{\tilde d_3^c} m_{\tilde e_3^c}}{m^4_{\tilde Q_3}m^2_{\tilde L_3}}
\label{sisy-3}
\end{eqnarray}
in Eq. (\ref{M_XM_Sigma-2}).  Here $m_{\tilde Q_i}$ stands for the mass of the first two family squark doublets, while $m_{\tilde Q_3}$ refers to the third family squark doublet mass, which are allowed to be different (and similarly for the other masses).

\begin{figure}[ht!]
\centering
\subfigure{\includegraphics[scale=1.1]{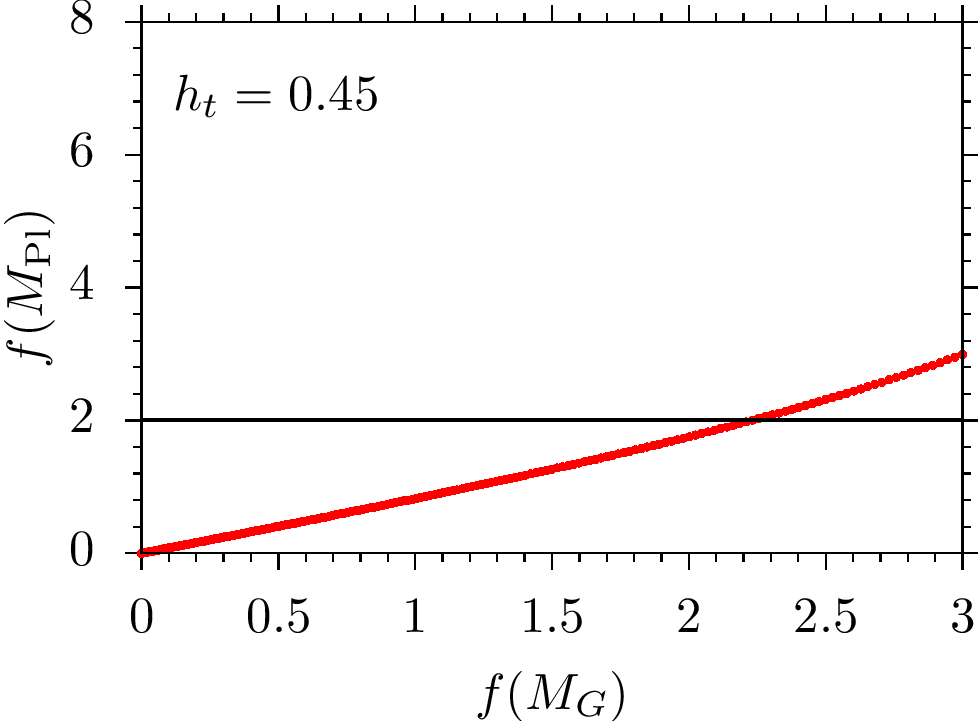}}%
\hspace*{0.18in}
\subfigure{\includegraphics[scale=1.1]{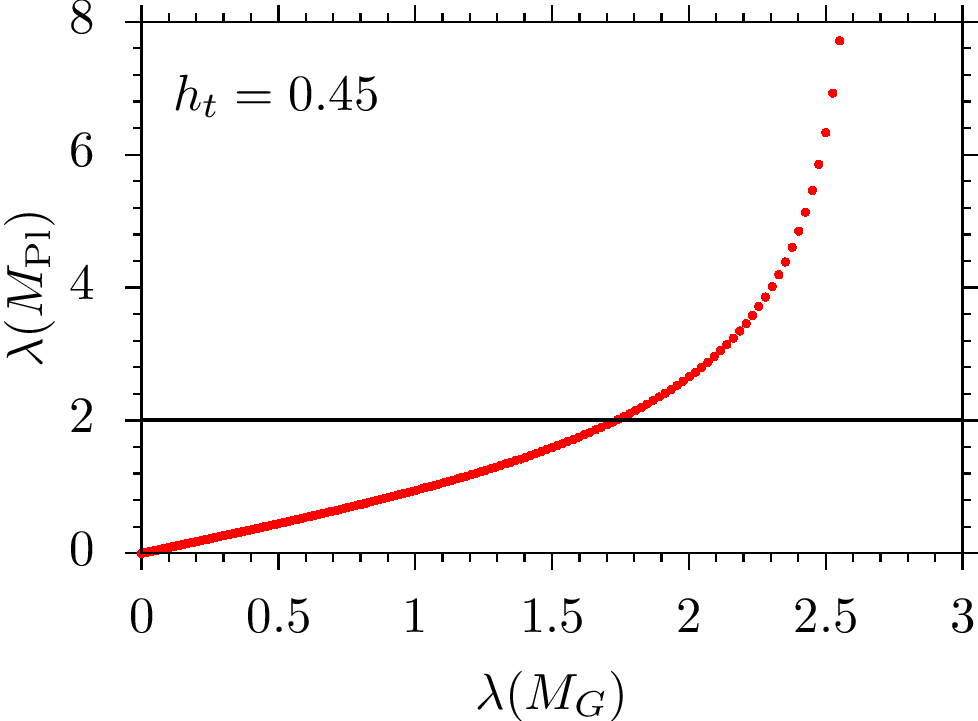}}
\caption{{Relation between the GUT and Planck scale values of $f$ (left) and $\lambda$ (right) defined in Eq. (\ref{gut poten}).}}
\label{fig-RGE4}
\end{figure}

We demand that the dimensionless couplings of the theory remain perturbative when extrapolated from the GUT scale to the Planck scale.  The minimal SUSY $SU(5)$ beta functions, for evolution above the $SU(5)$ scale, are given by \cite{Hisano:1992jj,Polonsky:1994sr}
\begin{eqnarray}
16 \pi^2 \frac{d g_5}{dt} &=& -3 g_5^3 \\
16 \pi^2 \frac{dh_t}{dt} &=& h_t \left(9 h_t^2 + 4 h_b^2 + \frac{12}{5}\lambda^2 -\frac{96}{5} g_5^2\right) \\
16\pi^2 \frac{dh_b}{dt} &=& h_b\left(10 h_b^2 + 3 h_t^2 + \frac{12}{5}\lambda^2-\frac{84}{5}g_5^2  \right)\\
16\pi^2 \frac{d \lambda}{dt} &=& \lambda \left( 3 h_t^2 + 4 h_b^2 + \frac{53}{10}\lambda^2 + \frac{21}{40} f^2 - \frac{98}{5} g_5^2 \right) \\
16\pi^2\frac{df}{dt} &=& f\left(\frac{63}{40}f^2 + \frac{3}{2}\lambda^2 - 30 g_5^2 \right)~.
\end{eqnarray}
Extrapolating these couplings from the GUT scale of $M_G = 2 \times 10^{16}$ GeV to the reduced Planck scale of $M_{Pl} = 2.4 \times 10^{18}$ GeV, with the GUT scale values $g_5 = 0.72$ and $h_t= 0.45$ (obtained from MSSM evolution of the low energy couplings), we obtain upper limits
\begin{equation}
f(M_{G}) \leq 2.24,~~~\lambda(M_G) \leq 1.75~,
\label{upper}
\end{equation}
for the GUT scale values of the superpotential couplings of Eq. (\ref{gut poten}). These limits are obtained by requiring that the couplings obey $|f(\mu)| \leq 2.0,\,|\lambda(\mu)| \leq 2.0$, which, we believe, are reasonable upper limits for the theory to remain perturbative.  The {evolutions} of the couplings $f$ and $\lambda$ above the GUT scale are shown in Fig. \ref{fig-RGE4}.
The constraints of Eq. (\ref{upper}) give upper bounds on the (common) mass $M_\Sigma$ of $\Sigma_8$ and $\Sigma_3$ fields and the mass $M_{H_C}$:
\begin{eqnarray}
M_\Sigma \leq 1.1 \,M_X,~~~~~M_{H_C} \leq 1.7 \,M_X~.
\label{octet-triplet 2}
\end{eqnarray}

\begin{figure}[ht!]
\centering
\subfigure{\includegraphics[scale=1.6]{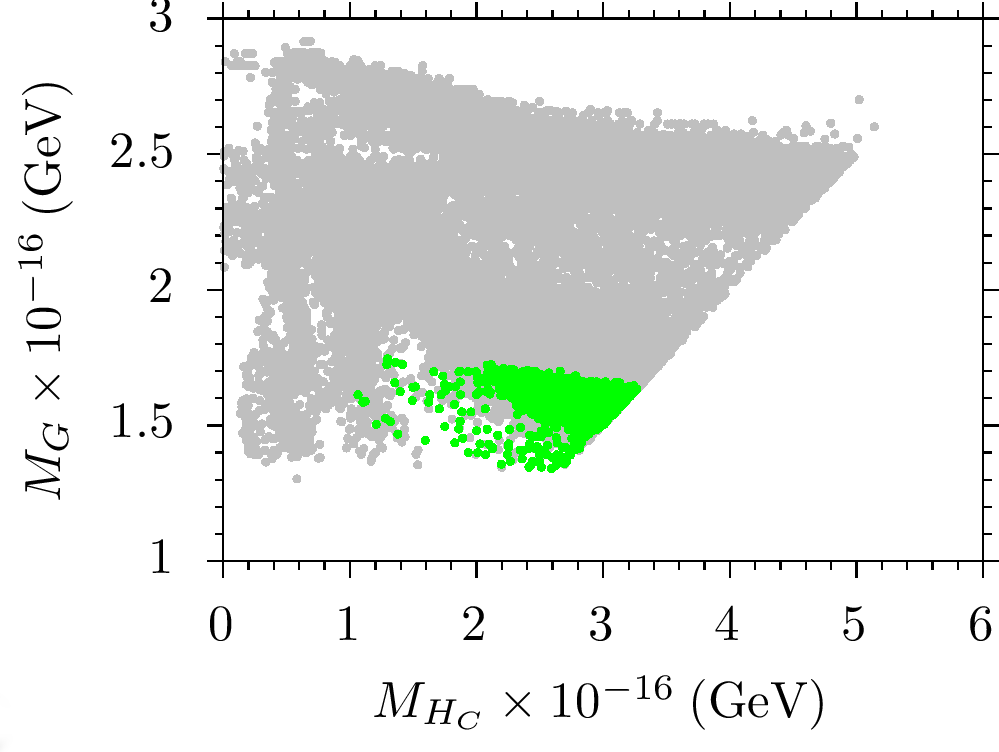}}%
\caption{Allowed parameter space in the $M_G -M_{H_{C}}$ plane in SUSY $SU(5)$ without threshold effects.  Here we use the definition  $M_G=(M_X^2M_{\Sigma})^{1/3}$. All points are compatible with the radiative electroweak symmetry breaking and the LSP being a neutralino. Green points are allowed by the Higgs boson mass, LHC  bounds on SUSY masses and $B$-physics constraints.}
\label{fig-GR2}
\end{figure}

Using low energy data and any specified  sparticle mass spectrum one can obtain values of the GUT scale mass parameters $M_G = ({M_X^2 M_\Sigma})^{\frac13}$ and $M_{H_C}$ from the relations given in Eqs. (\ref{eq:2.13})-(\ref{sisy-3}).  We have discussed the procedure to {calculate} the sparticle masses within our scenario in Sec. \ref{sec:scan}.  Adopting this procedure, we arrive at the ranges for the two GUT scale mass parameters $M_G$ and $M_{H_C}$. We have plotted these allowed ranges in Fig. \ref{fig-GR2}.  The grey shaded region satisfies radiative electroweak symmetry breaking constraints with a neutralino LSP. The green region, which is a subset of the grey region, also satisfies the Higgs boson mass constraint, $B$ meson decay limits, and lower limits on SUSY particles from the LHC searches.  From the grey region we find
  \begin{eqnarray}
  1.29 \times 10^{16}\, {\rm GeV} \lesssim  M_{G} \lesssim 2.92 \times 10^{16}  \,{\rm GeV},
        \label{bound0}\\
      0.43 \times 10^{14}\,  {\rm GeV} \lesssim M_{H_{C}} \lesssim 5.29 \times 10^{16} \, {\rm GeV},
      \label{bound1}
\end{eqnarray}
and from the green region we have
\begin{eqnarray}
1.34 \times 10^{16}\, {\rm GeV} \lesssim M_G \lesssim 1.78 \times 10^{16} \,{\rm GeV}, \label{boundonehalf}\\
1.0 \times 10^{16}\, {\rm GeV} \lesssim M_{H_{C}} \lesssim 3.2 \times 10^{16}\, {\rm GeV}~.
\label{bound2}
\end{eqnarray}
It is clear that the LHC bounds, especially from the  Higgs boson mass measurement and from the lower limit of about 2.1 TeV on the gluino mass, squeeze the allowed region in $M_{H_C}$ considerably (compare Eq. (\ref{bound1}) with Eq. (\ref{bound2})).
The ranges in these mass parameters arise primarily from the variations allowed in the SUSY mass parameters $m_{0_{1,2}}$, $m_{0_{3}}$ and $M_{1/2}$, with the uncertainties in the measurements of $\alpha_i(m_Z)$ playing a minor role. The quoted ranges in Eqs. (\ref{bound0})--(\ref{bound2}) correspond to choosing $m_{0_{1,2}},m_{0_{3}} \le 30$ TeV and $M_{1/2} \leq 2$ TeV.

We can combine Eqs. (\ref{boundonehalf}) and (\ref{bound2}) with Eq. (\ref{octet-triplet 2}) to obtain individual bounds in the LHC allowed region shown in green:
\begin{eqnarray}
M_X \geq 1.3 \times 10^{16} \,\rm{GeV},~~~
M_{\Sigma} \leq 1.90 \times 10^{16}\, \rm{GeV}~.
\label{m_Sigma-1}
\end{eqnarray}

Now we proceed to discuss Planck-suppressed operators and their influence on the ranges of the GUT scale masses derived here.  These modifications will have significant effects on the proton lifetime estimate that will be analyzed in Sec. \ref{sec:funparam}

\section{Planck-Suppressed Operators in SUSY \boldmath{$SU(5)$}}
\label{sec3}

In this section we discuss three types of Planck-suppressed operators that modify the GUT scale mass spectrum of SUSY $SU(5)$, as well as the light fermion masses, which have relevance to the $d=5$ proton decay estimates. These are (i) $d=5$ operators in the superpotential involving the light chiral multiplets of the MSSM, (ii) gravitational smearing of the gauge coupling unification arising from $d=5$ gauge kinetic corrections, and (iii) $d=5$ operators in the $SU(5)$-breaking superpotential. There are also possible modifications to the GUT scale boundary conditions on SUSY breaking parameters arising from Planck-suppressed operators; however, these corrections are effectively included in our analysis, since we allow for non-universal SUSY breaking parameters.
Analogous  discussions in SUSY $SO(10)$
can be found in Ref. \cite{Lucas:1996bc,Babu:1998wi,Babu:2010ej}.

\subsection{Correcting the wrong fermion mass relations}

While staying within minimal SUSY $SU(5)$, the wrong mass relations for the first two family fermions predicted by Eq. (\ref{mass}) can be corrected by allowing higher dimensional non-renormalizable operators in the Yukawa superpotential.\footnote{For attempts to correct the masses with soft SUSY breaking parameters see Ref. \cite{DiazCruz:2000mn,Enkhbat:2009jt}.}  Such operators will be suppressed by a fundamental scale, presumably the Planck scale.  The leading non-renormalizable operator that we include in our analysis is given by
\begin{equation}
W_{\rm Yuk}' = \sqrt{2} \kappa_{ij} \Psi_i^{ab} \Phi_{jc}\Sigma_b^c \overline{H}_a~.
\label{Yuknew}
\end{equation}
This corresponds to the $SU(5)$ contraction of the Higgs fields in the ${\bf 45^*}$ channel: {${\bf 24} \times {\bf 5^*} = {\bf 5^*}+{\bf 45^*}$}.  This operator will split the masses of the down-type quarks from those of charged leptons. The magnitude of the resulting Yukawa couplings are of order $\sigma/M_{\rm Pl} \sim 10^{-2}$, but even with such suppression, their contributions to $M_d$ and $M_{\ell}$ can be comparable to their respective experimental values. The $b$-quark to $\tau$-lepton mass ratio which is close to 1 at the GUT scale does show deviation from 1 at the level of 25\%.  The Planck-induced terms of Eq. (\ref{Yuknew}) can also correct this relation, provided that $\tan\beta$ is not too large (or else the strength of the relevant operator would  be insufficient for a 25\% correction).
While in principle one could also write operators in the ${\bf 5^*}$ channel in Eq. (\ref{Yuknew}), we do not include them as this contraction will not split $M_d$ from $M_\ell$. Similarly, allowed higher dimensional operators of the type $\epsilon_{abcfe} \Psi_i^{ab} \Psi_j^{cd}\Sigma^f_d H^e$ are not included, as they are not necessary to correct light fermion masses.

It is interesting to note that  the higher dimensional Yukawa operators of Eq. (\ref{Yuknew}) can be generated by integrating out a ${\bf 5}+ {\bf 5^*}$ matter fields, {as a simplest example},  with mass of order the Planck scale. Denoting these fields as $\chi + \overline{\chi}$, the superpotential given by \cite{Babu:2012pb}
\begin{equation}
W(\chi+\overline{\chi}) = M_\chi \chi \overline{\chi} + \Psi \overline{\chi} \overline{H} + \Phi \chi \Sigma
\end{equation}
will generate terms of Eq. (\ref{Yuknew}), without inducing other terms.
We make one simplifying assumption in our analysis.  We assume that the terms of
Eq. (\ref{Yuknew}) are diagonal in flavor space in a basis where $f^{ij}$ of Eq. (\ref{Yukexp}) is also diagonal.  All quark mixings will then arise from the $h^{ij}$ couplings of Eq. (\ref{Yukexp}). {Deviations from this assumption would result in order one corrections to the $d=5$ proton decay rate that we estimate in Sec. \ref{sec:funparam}. }
With this assumption, including Eq. (\ref{Yuknew}), the mass matrices of down-type quarks and charged leptons take the form:
\begin{eqnarray}
M_d = (f+f')v_d,~~~M_\ell = \left(f-\frac{3}{2} f'\right) v_d~.
\label{massnew}
\end{eqnarray}
Here we have defined a dimensionless coupling
\begin{equation}
f' = 2 \kappa \sigma
\end{equation}
which has elments at most of order few $\times 10^{-2}$, {where $\kappa$ has inverse mass dimension as defined in Eq. (\ref{Yuknew}).} The six free Yukawa couplings $f_i$ and $f_i'$ of the two diagonal matrices can now be used to fit consistently down-quark and charged lepton masses. Thus, this modification rectifies the wrong mass relations of minimal renormalizable $SU(5)$ in a simple way.

The addition of Eq. (\ref{Yuknew}) into the Yukawa superpotential modifies the color-triplet Higgs couplings to the fermions.  Noting the relations in Eq. (\ref{massnew}), the color-triplet Higgs couplings to fermions now become
\begin{eqnarray}
W(H_C,\overline{H}_C) &=& -\frac{1}{2}h^{ij} Q_i^{\alpha p} Q_j^{\beta q} H_C^\gamma \epsilon_{\alpha \beta \gamma} \epsilon_{pq} + \left(\frac{M_{\ell}}{v_d}\right)_i Q_i^{\alpha p}L_i^q \overline{H}_{C\alpha} \epsilon_{pq} \nonumber \\ &+& h^{ij} u^c_{i \alpha} e^c_j H_C^\alpha -\left(\frac{M_d}{v_d}\right)_iu^c_{i\alpha} d^c_{i \beta} \overline{H}_{C\gamma} \epsilon^{\alpha \beta \gamma}~.
\end{eqnarray}
We can write down these interactions in the mass eigenbasis of the quarks and leptons.  Since $M_d$ and $M_\ell$ are chosen to be diagonal, we simply have to absorb any phases in these two sectors, and use the relation
\begin{eqnarray}
h = \hat{V}^T \left(\frac{M_u}{v_u} \right) \hat{V}
\end{eqnarray}
where
\begin{equation}
\hat{V} = P V Q
\end{equation}
with $V$ being the CKM matrix in the standard phase convention, and $P$, $Q$ being diagonal phase matrices.  The phases in $Q$ can be absorbed into fermion fields, but those in $P$ will remain in the color-triplet Higgs couplings.  We denote $P_i = e^{i \phi_i/2}$, with the condition
\begin{equation}
\phi_1 + \phi_2 + \phi_3 = 0~.
\end{equation}

We can now write the effective $d=5$ baryon number violating operators by integrating out the color-triplet Higgs(ino) fields \cite{Nath:1988tx,Hisano:1992jj}:
\begin{eqnarray}
W_{d=5} &=& \frac{1}{2 M_{H_C}}\left(\frac{M_u}{v_u}  \right)_ie^{i \phi_i}V_{kl}^*\left(\frac{M_\ell}{v_d}  \right)_l(Q_iQ_i)(Q_kL_l)
 \nonumber \\
&+&
\frac{1}{M_{H_C}}\left(\frac{M_u}{v_u}  \right)_i V_{ij} e^{-i\phi_k} V_{kl}^* \left(\frac{M_d}{v_d}  \right)_l(u_i^c e_j^c)(u_k^c d_l^c)~.
\label{baryo}
\end{eqnarray}
Here the contractions are defined as follows:
\begin{eqnarray}
(Q_i Q_j)(Q_k L_l) &=& \epsilon_{\alpha \beta \gamma} (u_i^\alpha d_j^{\prime \beta}-d_i^{\prime \alpha} u_j^\beta)(u_k^\gamma e_l-d^{\prime\gamma}_k \nu_l) \nonumber \\
(u^c_i e^c_j)(u^c_k d^c_l) &=& \epsilon^{\alpha \beta \gamma} u^c_{i \alpha}e^c_j u^c_{k \beta} d^c_{l \gamma}~
\end{eqnarray}
with $d' = V.d$.

The form of Eq. (\ref{baryo}) is identical to the one studied in renormalizable SUSY $SU(5)$, but the $LLLL$ operator has charged lepton masses, rather than down quark masses, and in the $RRRR$ operator it is the down quark masses that appear. Since we accommodate the mass ratio $m_\mu^0/m_s^0 \simeq 4.4$ consistently here, the $d=5$ proton decay rate becomes enhanced by a factor of $(4.4)^2 \simeq 20$, which sets more severe constraints on the model parameters compared to the case when $m_\mu^0/m_s^0 =1$ is used.  It is this form of the effective baryon number violating operators that we shall use in our numerical study.


\subsection{Gravitational smearing of unified gauge coupling}

\begin{figure}[ht!]
\centering
\subfigure{\includegraphics[scale=1.1]{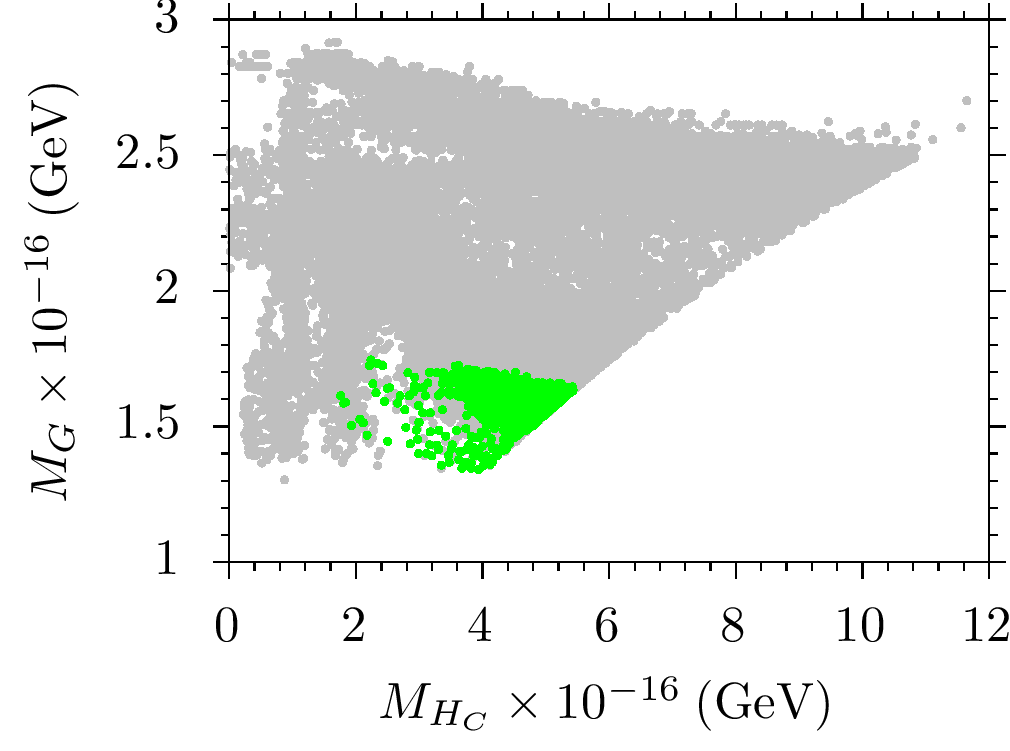}}%
\subfigure{\includegraphics[scale=1.1]{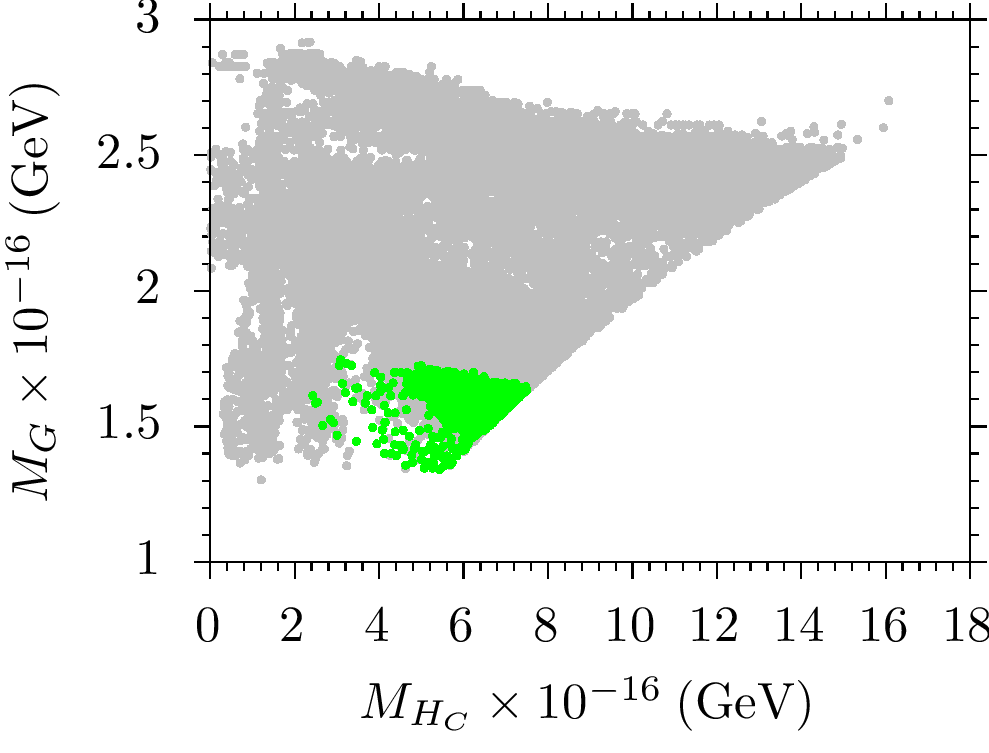}}
\caption{Allowed parameter space in the $M_G -M_{H_{C}}$ plane in SUSY $SU(5)$ with threshold effects causing smearing of the unified gauge coupling (left panel) and from the $\Sigma^4$ superpotential  corrections (right panel).
All points are compatible with the REWSB and the LSP being a neutralino. Green points show the allowed region from the Higgs boson mass, $B$-physics constraints and LHC limits on SUSY masses.}
\label{fig-GR3}
\end{figure}

In presence of quantum gravity, the gauge boson kinetic terms could receive corrections through the effective $d=5$ Lagrangian given as \cite{Hill:1983xh,Shafi:1983gz,Hall:1992kq,Dasgupta:1995js}
\begin{equation}
    \delta {\cal L} = \frac{c}{2M_{\rm Pl}}{\rm tr}(G_{\mu\nu} G^{\mu\nu} \Sigma)
    \label{smear}
\end{equation}
where $G^{\mu\nu} = G^{\mu\nu}_aT^a$ is the $SU(5)$ field strength with the generators normalized as ${\rm Tr}(T^a T^b) = \frac{1}{2}\delta^{ab}$.  The unified gauge coupling will be smeared in presence of Eq. (\ref{smear}) with the $\alpha_5^{-1}$ in the solution to the one-loop for the gauge couplings replaced by
\begin{equation}
\alpha_5^{-1} \rightarrow \alpha_5^{-1} + \frac{c \,\sigma \alpha_5^{-1}}{2 M_{\rm Pl}} \left(
 \begin{matrix}
 -1 \\
 -3 \\
 +2\\
 \end{matrix}
 \right)
\end{equation}
where the three entries correspond to the smearing of  $\alpha^{-1}_{1,2,3}$ in that order.  As a result, Eq. (\ref{M_XM_Sigma-2}) will be modified to
\begin{equation}
(- 2 \alpha_3^{-1} + 3 \alpha_2^{-1} - \alpha_1^{-1} ) (m_Z)
	= \frac{1}{2\pi} \left\{
		\frac{12}{5} \, \ln \frac{M_{H_C}}{m_Z}
		-2 \ln \frac{m_{SUSY}}{m_Z} \right\} - \frac{6 \sigma \,c}{ M_{\rm Pl}}\alpha_5^{-1}		\label{M_XM_Sigma-3}~.
\end{equation}
This threshold correction does not however modify Eq. (\ref{M_XM_Sigma-1}).  The allowed parameter space of the model is plotted in Fig. \ref{fig-GR3}, left panel in the $M_{H_C}-M_G$ plane,  corresponding to the choice $c=1$. This should be compared with the allowed region without threshold effects shown in Fig. \ref{fig-GR2}. From the grey region of this figure we obtain the allowed range (for $c=1$)
\begin{eqnarray}
0.81 \times 10^{14} \,{\rm GeV}\lesssim M_{H_{C}} \lesssim 1.16 \times 10^{17} \,{\rm GeV},
\end{eqnarray}
{while the green region corresponds to}
\begin{eqnarray}
1.88 \times 10^{16}\, {\rm GeV}\lesssim M_{H_{C}} \lesssim 5.25 \times 10^{16}\, {\rm GeV}~.
\end{eqnarray}
The value of $M_G$ does {is not} altered from this threshold effect. As far as the modification of Eq. (\ref{M_XM_Sigma-2}) as shown in Eq. (\ref{M_XM_Sigma-3}) is concerned, it can be interpreted as redefinitions of the mass parameters of Eq. (\ref{M_XM_Sigma-2}) such that values of the mass parameters are multiplied with exponential factors as follows:

\begin{eqnarray}
M_{H_C} &\rightarrow& M_{H_C} \times {\rm exp}{\left(\frac{-5\pi\sigma c}{\alpha_5 M_{\rm Pl}}\right)}, \\
M_{X} &\rightarrow& M_{X} \times {\rm exp}{\left(\frac{3\pi\sigma c}{10\alpha_5 M_{\rm Pl}}\right)},  \\
M_{\Sigma} &\rightarrow& M_{\Sigma} \times {\rm exp}{\left(\frac{-3\pi\sigma c}{5\alpha_5 M_{\rm Pl}}\right)}~.
\end{eqnarray}
It should be noted that these are not the physical masses of particles, but rather are effective masses which would capture the effects of included threshold corrections.

\subsection{Corrections to the symmetry breaking superpotential}

The superpotential of Eq. (\ref{gut poten}) can receive Planck-suppressed correction of dimension five:
\begin{equation}
W' = \frac{\kappa_1}{4} {\rm tr}(\Sigma^4) + \frac{\kappa_2}{4} \left({\rm tr}(\Sigma^2)\right)^2~
\end{equation}
where $\kappa_{1,2}$ have inverse dimensions of mass.  Including these terms in the symmetry breaking analysis shows that the color-octet $\Sigma_8$ and the $SU(2)_L$-triplet $\Sigma_3$ are no longer degenerate, with their masses given by
\begin{eqnarray}
M_{\Sigma_8} &=& \frac{5}{2} f\sigma + \frac{5}{2}\kappa_1 \sigma^2 \\
M_{\Sigma_3} &=& \frac{5}{2} f \sigma - 10 \kappa_1 \sigma^2~.
\label{Sigma3}
\end{eqnarray}
(Since the physical masses are defined to be positive, we have flipped the sign of $M_{\Sigma_3}$ in Eq. (\ref{Sigma3}).)
In presence of this mass splitting, the relation {in} Eq. (\ref{M_XM_Sigma-2}) will be modified to
\begin{eqnarray}
(- 2 \alpha_3^{-1} + 3 \alpha_2^{-1} - \alpha_1^{-1} ) (m_Z)
	= \frac{1}{2\pi} \left\{
		\frac{12}{5} \, \ln \frac{M_{H_C}}{m_Z} + 6\, {\rm ln} \frac{M_{\Sigma_8}}{M_{\Sigma_3}}
		-2 \ln \frac{m_{SUSY}}{m_Z} \right\}.
		\label{M_XM_Sigma-4}
\end{eqnarray}
And the relation in Eq. (\ref{M_XM_Sigma-1}) will be modified to
\begin{eqnarray}
(- 2 \alpha_3^{-1} - 3 \alpha_2^{-1} + 5 \alpha_1^{-1} ) (m_Z)
	&=& \frac{1}{2\pi} \left\{
		12 \, \ln \frac{M_X^2 M_\Sigma}{m_Z^3}
		+ 8 \ln \frac{m_{SUSY}}{m_Z}-6\,{\rm ln}\frac{M_{\Sigma_8}}{M_{\Sigma_3}} \right\},
		\label{M_XM_Sigma-5}
\end{eqnarray}
Here $M_\Sigma= (5/2) f \sigma$ is defined to be the (common) mass of the color octet and weak triplet from $\Sigma$.

We have plotted the allowed parameter space of the model including these threshold corrections in the  $M_{H_C}-M_G$ plane in Fig. \ref{fig-GR3} on the right panel, with the assumption that the shift in mass of $H_C$ is up to about 40\%.  From here we find that in the gray region we have the $H_C$ mass range give by
\begin{eqnarray}
0.81 \times 10^{14} \, {\rm GeV} < M_{H_{C}} < 1.52 \times 10^{17}\, {\rm GeV},
\end{eqnarray}
while in the green region the range is
\begin{eqnarray}
0.43 \times 10^{16} \, {\rm GeV}<M_{H_{C}} < 7.2 \times 10^{16}\, {\rm GeV}
\end{eqnarray}

\begin{figure}[ht!]
\centering
\subfigure{\includegraphics[scale=1.5]{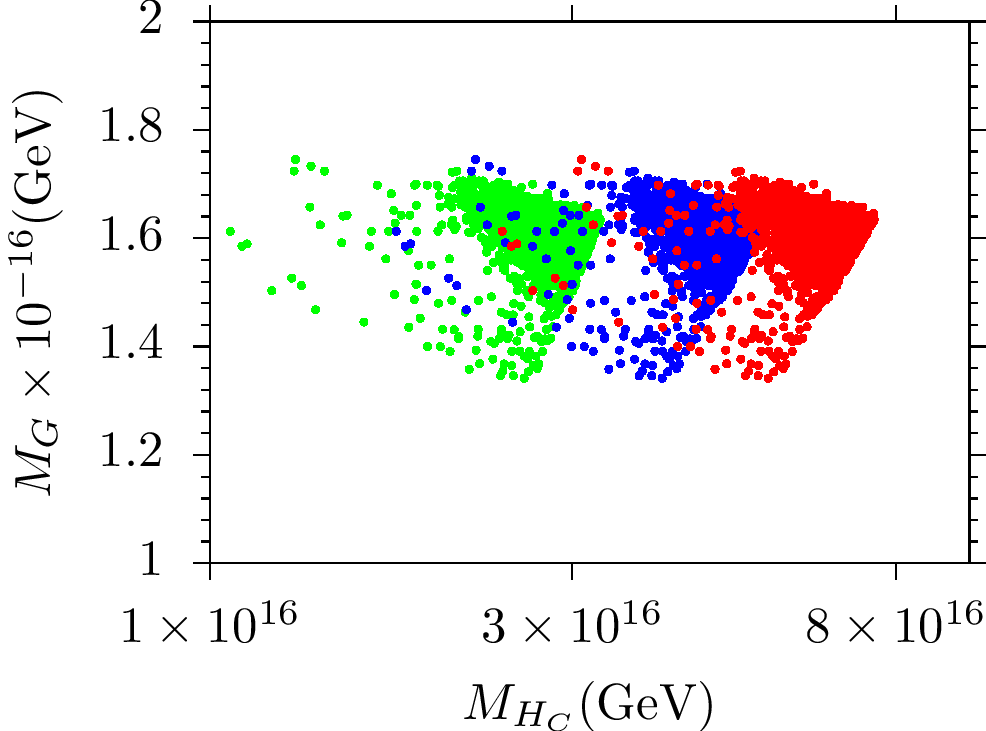}}
\caption{Allowed parameter space in the $M_{H_C}-M_G$ plane with the two types of threshold effects discussed in this section.  The red shaded region arises from the superpotential corrections and allow the largest effective $H_C$ mass.  The blue shaded region is from gravitational smearing of the unified gauge coupling and the green shaded region is without GUT scale threshold effects, Here REWSB constraint, neutralino LSP condition, Higgs boson mass, $B$-physics constraints and LHC limits on SUSY particle masses are all satisfied.}
\label{fig-RGE444}
\end{figure}

{We have summarized the results for the allowed region of parameters in the $M_{H_C}-M_G$ plane including the threshold effects in Fig. \ref{fig-RGE444}.}  The red shaded region, corresponding to the superpotential correction, is seen to increase the effective mass of the color-triplet Higgs to values as large as {about} $7 \times 10^{16}$ GeV.  In principle, the various corrections could  act collectively, which could further increase the $H_C$ mass. We shall however not assume this, and in our numerical analysis on proton decay, we fix $M_{H_C} = 7 \times 10^{16}$ GeV as an effective mass.

It should be noted that there could be other types of Planck-suppressed operators that could contribute to $d=5$ proton decay amplitude. For example, ${\bf 10}^i\, {\bf 10}^j\, {\bf 10}^k\, \overline{{\bf 5}}_\ell$ terms, involving the fermion superfields and suppressed by one power of the Planck mass, could be present.  The coefficients of such operators should be of order $10^{-7}$ or smaller, in order to be consistent with proton lifetime limits.  These operators do not help in correcting the wrong mass relations for the light quark and leptons.  We have not included such operators in our study.

\section{Scanning Procedure, Parameter Space and Experimental Constraints}
\label{sec:scan}

This section summarizes the scanning procedure and constraints which we apply in our analyses.  We have performed random scans in the fundamental parameter space as follows:

\begin{equation}
\setstretch{1.2}
\begin{array}{ccc}
0  \leq & m_{0_{1,2}}, m_{0_{3}} & \leq 30 ~{\rm TeV}~, \\
0  \leq & M_{1/2} & \leq 2 ~{\rm TeV}~, \\
1.2 \leq & \tan\beta & \leq 20~, \\
-3 \leq & A_{0}/m_{0_{3}} & \leq 3~, \\
-10~{\rm TeV}  \leq & A_{0} & \leq 10 ~{\rm TeV}~,\\
0  \leq & \mu,m_{A} & \leq 30 ~{\rm TeV}~.
\end{array}
\label{paramSP}
\end{equation}
Here  $m_{0_{1,2}}$  and $m_{0_{3}}$ are  the universal SSB {masses} for MSSM first two  and third family sfermions {respectively}. This choice of split masses for the sfermions is motivated by flavor symmetry as discussed in the context of sMSSM \cite{Babu:2014sga,Babu:2014lwa}. $M_{1/2}$ is the gaugino mass parameter, $\tan\beta\equiv v_{u}/v_{d}$ represents the ratio of the  VEVs of the MSSM Higgs doublets $H_{u}$ and $H_{d}$. $A_{0}$ is the universal SSB trilinear scalar interaction (with corresponding Yukawa couplings factored out). The parameters $\mu$ and $m_{A}$ are the Higgs bilinear mass term and the mass of the CP-odd Higgs boson respectively.

We implement the randomly determined boundary conditions to ISAJET 7.84 package~\cite{Paige:2003mg} , which calculates the mass spectrum for the supersymmetric particles and Higgs bosons. In addition to the mass spectrum, ISAJET is interfaced with IsaTools package \cite{Baer:2002fv} to calculate rare $B-$meson decays such as $B_{s}\rightarrow \mu^{+}\mu^{-}$ and $B\rightarrow X_{s}\gamma$ and $B_{u}\rightarrow \tau\nu$ as well as the dark matter observables. ISAJET uses the weak scale values of the SM gauge couplings and the third family Yukawa couplings, which are evolved to gauge coupling unification scale $M_{{\rm U}}$ through MSSM RGEs imposed in $\overline{DR}$ regularization scheme. The gauge coupling unification scale $M_{{U}}$ is determined numerically as the scale at which the RG evolution of the gauge couplings coincide each other.\footnote{We denote by $M_U$ the scale where the two gauge couplings $\alpha_1$ and $\alpha_2$ unify in the absence of any threshold corrections. While related, $M_U$ is not quite the same as $M_G$, which also includes threshold effects.}
However, in the evolution of the gauge couplings, the unification condition is not applied strictly, since a few percent deviation from unification can be assigned to unknown GUT-scale threshold corrections~\cite{Hisano:1992jj,Yamada:1992kv,Chkareuli:1998wi}. { The deviation $g_{3}$ from $g_{1}=g_{2}$ at $M_{{\rm U}}$ is about a few percent.} In addition to the gauge and Yukawa couplings, we set $m_{t}=173.3$ GeV as the central value of top quark mass \cite{Group:2009ad}. Note that  $1-2\sigma$ variation in the top quark mass can result in 1-2 GeV difference in the Higgs boson mass \cite{Ajaib:2013zha}.

The various boundary conditions are imposed at $M_{\rm U}$ and all the SSB parameters, along with the gauge and Yukawa couplings, are evolved back to the weak scale including the SUSY threshold corrections \cite{Pierce:1996zz}. The entire parameter set is iteratively run between $M_{\rm Z}$ and $M_{\rm U}$  using the full 2-loop RGEs until a stable {solution is}  obtained.

One of the important constraint comes from the cosmological abundance of the charged particles \cite{Nakamura:2010zzi}, which prevents them to be stable and excludes the regions in the parameter space where a charged particle happens to be the lightest supersymmetric particle (LSP). In this context, we accept only the solutions for which one of the neutralinos is the LSP and it is accounted for saturating the relic density of dark matter. In addition to the cosmological constraints, we also require all the solutions to satisfy requirement of radiative electroweak symmetry breaking (REWSB). After generating the data consistent with these conditions, it is subsequently subjected to the  mass bounds on  the particles \cite{Nakamura:2010zzi} including the Higgs boson \cite{Aad:2012tfa,Chatrchyan:2012ufa} and the gluino \cite{Aaboud:2017vwy}, the constraints from the rare $B-$meson decays such as $B_{s}\rightarrow \mu^{+}\mu^{-}$ \cite{Aaij:2012nna}, $B_{s}\rightarrow X_{s}\gamma$ \cite{Amhis:2012bh}, and $B_{u}\rightarrow \tau\nu_{\tau}$ \cite{Asner:2010qj}. We also include WMAP measurements on the dark matter relic density \cite{Hinshaw:2012aka}. Eq.(\ref{constraints}) summarizes the constraints successively applied to the data in our analyses:

\begin{equation}
\setstretch{1.8}
\begin{array}{c}
m_h  = 123-127~{\rm GeV}
\\
m_{\tilde{g}} \geq 2.1~{\rm TeV}
\\
0.8\times 10^{-9} \leq{\rm BR}(B_s \rightarrow \mu^+ \mu^-)
  \leq 6.2 \times10^{-9} \;(2\sigma)
\\
2.99 \times 10^{-4} \leq
  {\rm BR}(B \rightarrow X_{s} \gamma)
  \leq 3.87 \times 10^{-4} \; (2\sigma)
\\
0.15 \leq \dfrac{
 {\rm BR}(B_u\rightarrow\tau \nu_{\tau})_{\rm MSSM}}
 {{\rm BR}(B_u\rightarrow \tau \nu_{\tau})_{\rm SM}}
        \leq 2.41 \; (3\sigma) \\
   0.0913 \leq \Omega_{{\rm CDM}}h^{2} \leq 0.1363
\label{constraints}
\end{array}
\end{equation}

Before concluding this discussion, we should note that the latest release from the Planck Satellite on the DM relic density measurements \cite{Akrami:2018vks} provides more restrictive bound on the relic abundance of the LSP neutralino as $0.114 \leq \Omega h^{2}\leq 0.126~(5\sigma)$. Considering the large uncertainties in calculation of the relic abundance arising from non-linearity of the Boltzmann equation and its exponential solutions we employ the less restrictive  WMAP bound in our analyses.


\section{Proton Decay and Fundamental Parameter Space of SUSY \boldmath{$SU(5)$}}
\label{sec:funparam}

\begin{figure}[h!]
\centering
\subfigure{\includegraphics[scale=1.1]{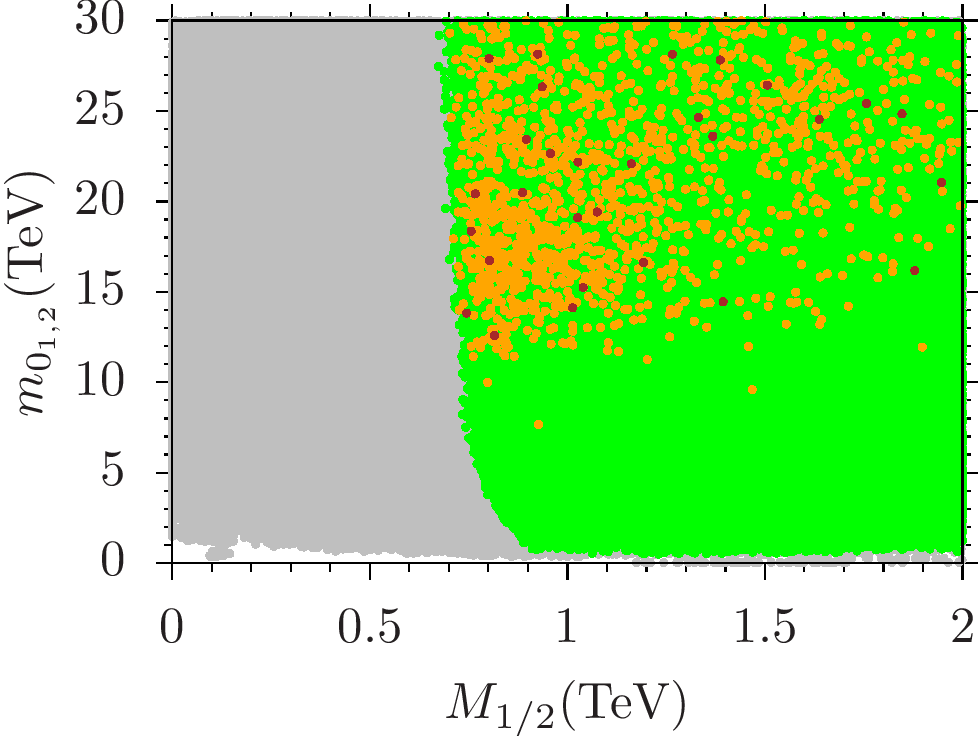}}%
\subfigure{\includegraphics[scale=1.1]{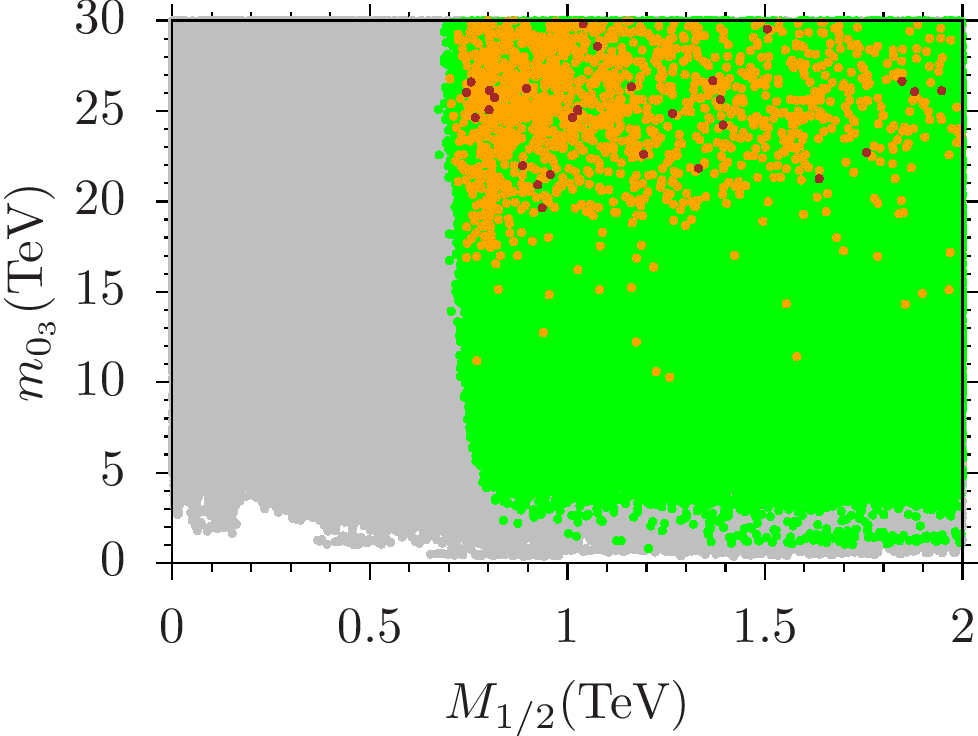}}
\subfigure{\includegraphics[scale=1.1]{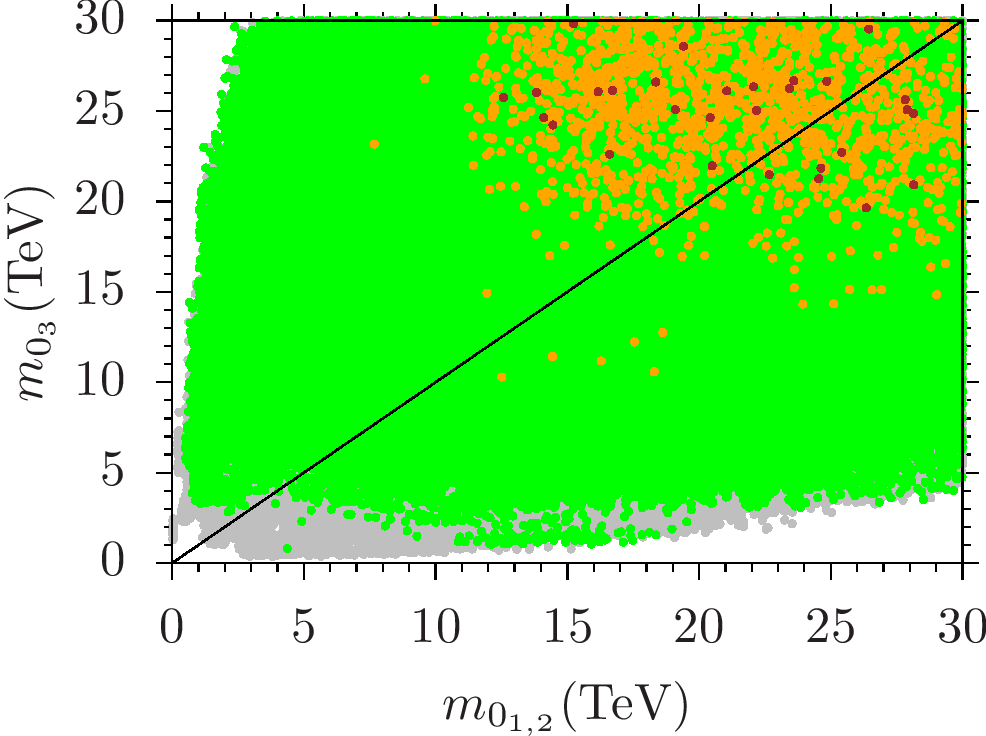}}%
\subfigure{\includegraphics[scale=1.1]{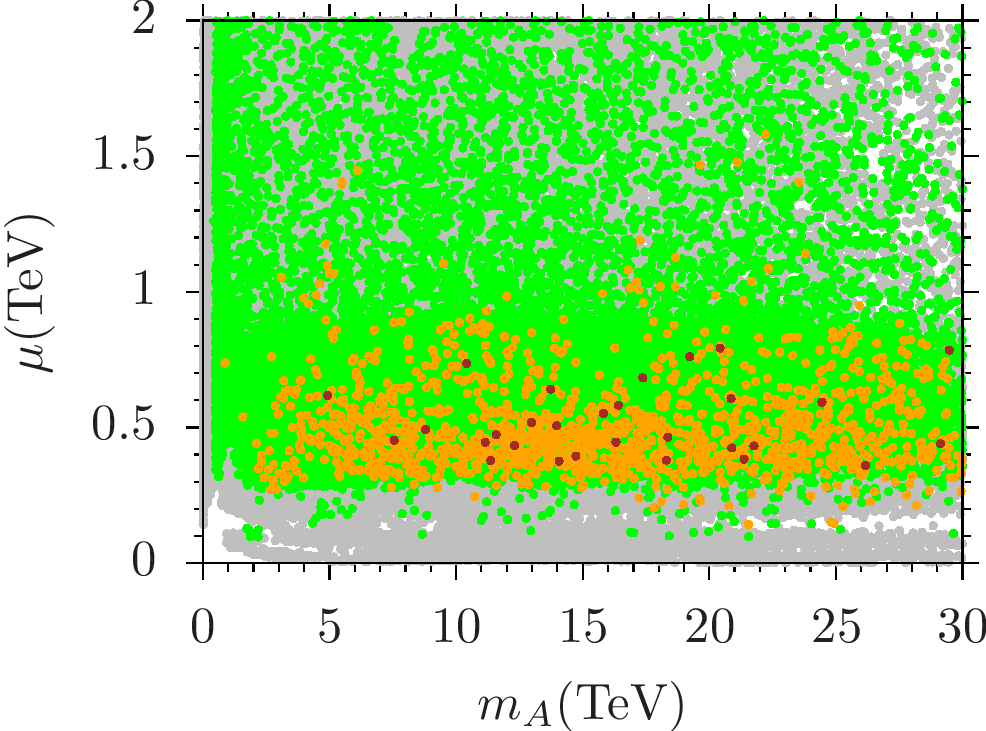}}
\caption{Allowed parameter space in the $m_{0_{1,2}}-M_{1/2}$, $m_{0_{3}}-M_{1/2}$, $m_{0_{3}}-m_{0_{1,2}}$ and $\mu-m_{A}$ plane. All points are compatible with the REWSB and neutralino LSP conditions. Green points represent the solution which are consistent with the mass bounds {on sparticles and Higgs boson}, and constraints from rare $B-$meson decays. The regions consistent with the current proton lifetime are shown in orange as a subspace of green. The bounds from the WMAP measurements are applied on top of the proton lifetime constraint, and the consistent solutions are shown in brown. { The diagonal line in the $m_{0_{3}}-m_{0_{1,2}}$ plane shows the region where $m_{0_{3}}=m_{0_{1,2}}$}.}
\label{fig1}
\end{figure}

In this section, we discuss the fundamental parameter space of the SUSY $SU(5)$ model with supersymmetry breaking parametrized by Eq. (\ref{paramSP}) and identify the mass spectrum compatible with the bound on the proton lifetime. Fig. \ref{fig1} displays allowed parameter space in the $m_{0_{1,2}}-M_{1/2}$, $m_{0_{3}}-M_{1/2}$, $m_{0_{3}}-m_{0_{1,2}}$ and $\mu-m_{A}$ planes. All points are compatible with the REWSB and neutralino LSP conditions. Green points represent the solutions which are consistent with the mass bounds {on sparticles and Higgs boson}, and constraints from rare $B-$meson decays. The regions consistent with the current proton lifetime are shown in orange as a subspace of green. The bounds from the WMAP measurements are applied on top of the proton lifetime constraint, and the consistent solutions are shown in brown. { The diagonal line in the $m_{0_{3}}-m_{0_{1,2}}$ plane shows the region where $m_{0_{3}}=m_{0_{1,2}}$}. The proton lifetime is calculated by setting the triplet Higgsino mass to be $M_{H_C} = 7 \times 10^{16}$ GeV. The $m_{0_{1,2}}-M_{1/2}$ plane shows that any value {greater than 5 TeV for the SSB mass term for the first two families can be compatible with the bound on the proton lifetime, while the relic density constraint raises the bound on $m_{0_{1,2}}$ up to about 10 TeV. On the other hand}, the regions with $M_{1/2}\lesssim 700$ GeV (gray region) are excluded. This exclusion  arises due to the gluino mass limits.  One can see a stronger impact from the proton lifetime on $m_{0_{3}}$ from the $m_{0_{3}}-M_{1/2}$ plane. {The  orange points which are  compatible with the bound on the proton lifetime are mostly accumulated in the regions with $m_{0_{3}} \gtrsim 10$ TeV.} We also present our findings in the $m_{0_{3}}-m_{0_{1,2}}$ plane. Here  $m_{0_{3}}$ can take relatively smaller values when $m_{0_{1,2}} \gtrsim 15$ TeV, which leads to quite heavy spectrum for the first two family  sfermions. { The diagonal line corresponds to the limits when we have universal SSB mass terms for all squarks and sleptons.}
The $\mu-m_{A}$ plane shows that solutions in orange can be realized only when $\mu \lesssim 1.5$ TeV, even though it is varied up to 30 TeV in our scan. The dark matter relic density condition constrains $\mu$ further as $\mu \lesssim 1$ TeV (brown points). The regions with low $\mu$ might be favored by the fine-tuning arguments \cite{Baer:2012up}  and also they can provide interesting DM predictions which can be tested in direct detection experiments.

\begin{figure}[ht!]
\centering
\subfigure{\includegraphics[scale=1.1]{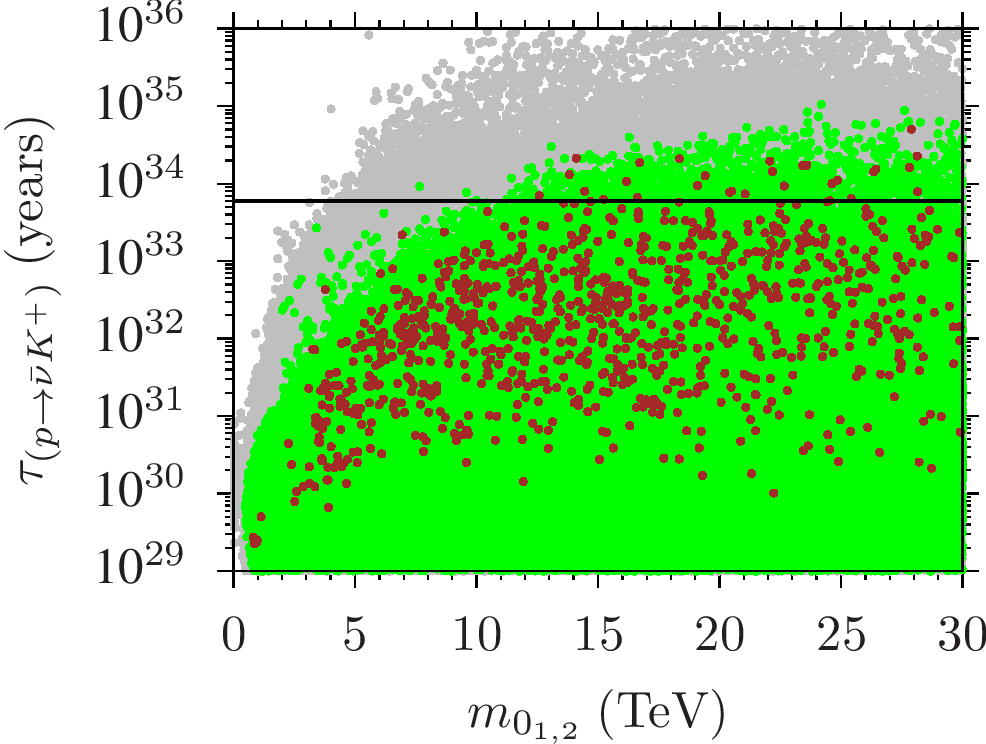}}%
\subfigure{\includegraphics[scale=1.1]{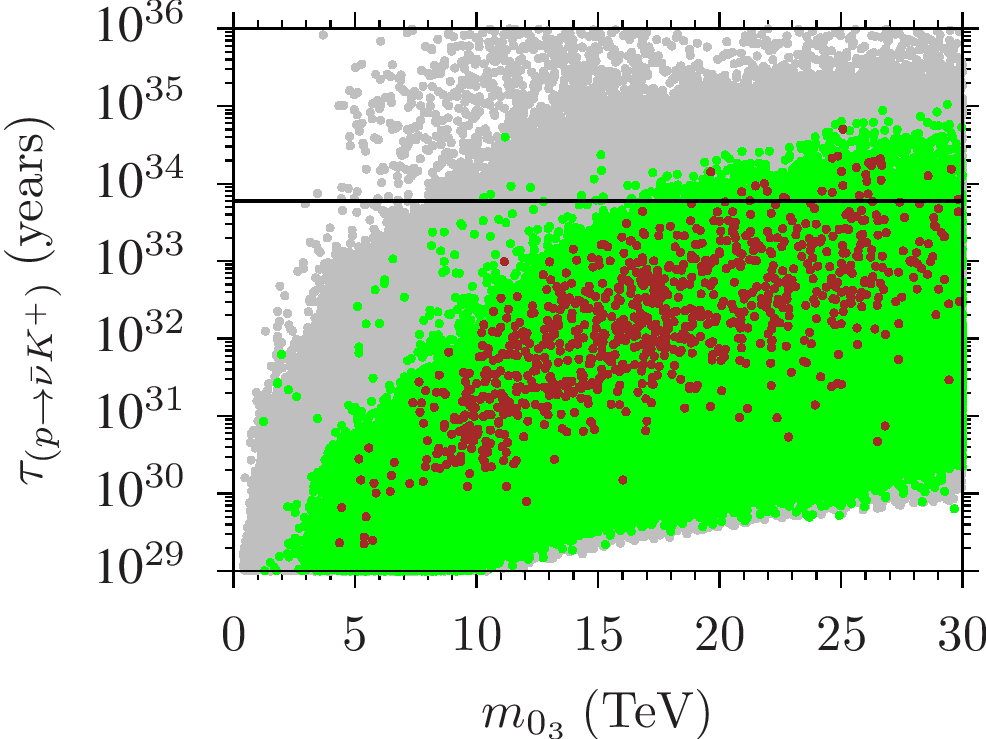}}
\subfigure{\includegraphics[scale=1.1]{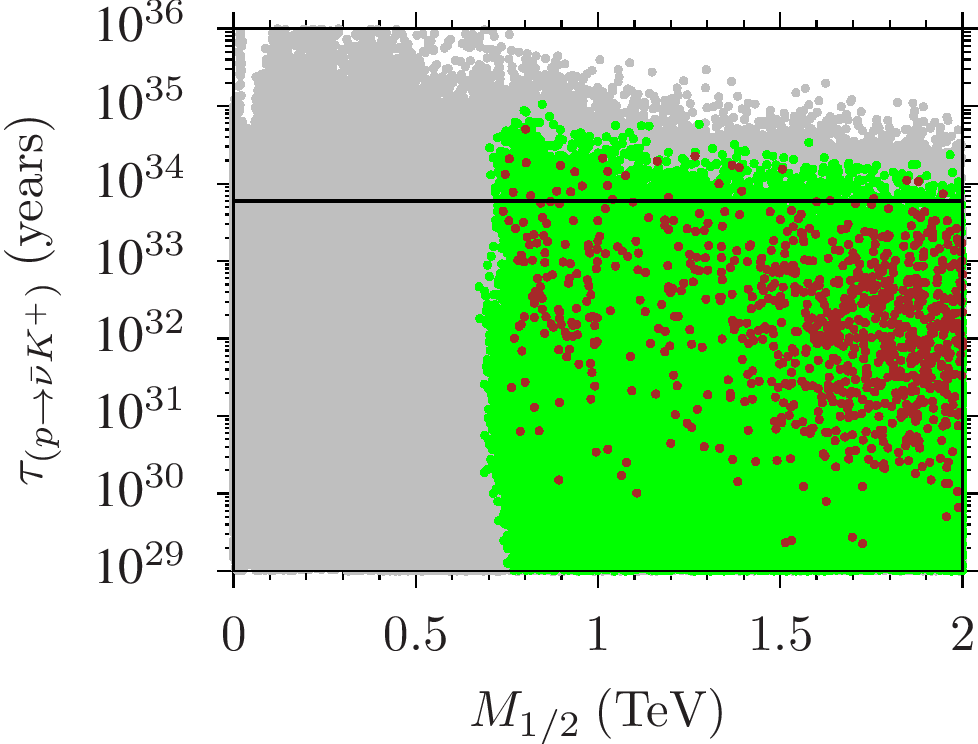}}%
\subfigure{\includegraphics[scale=1.1]{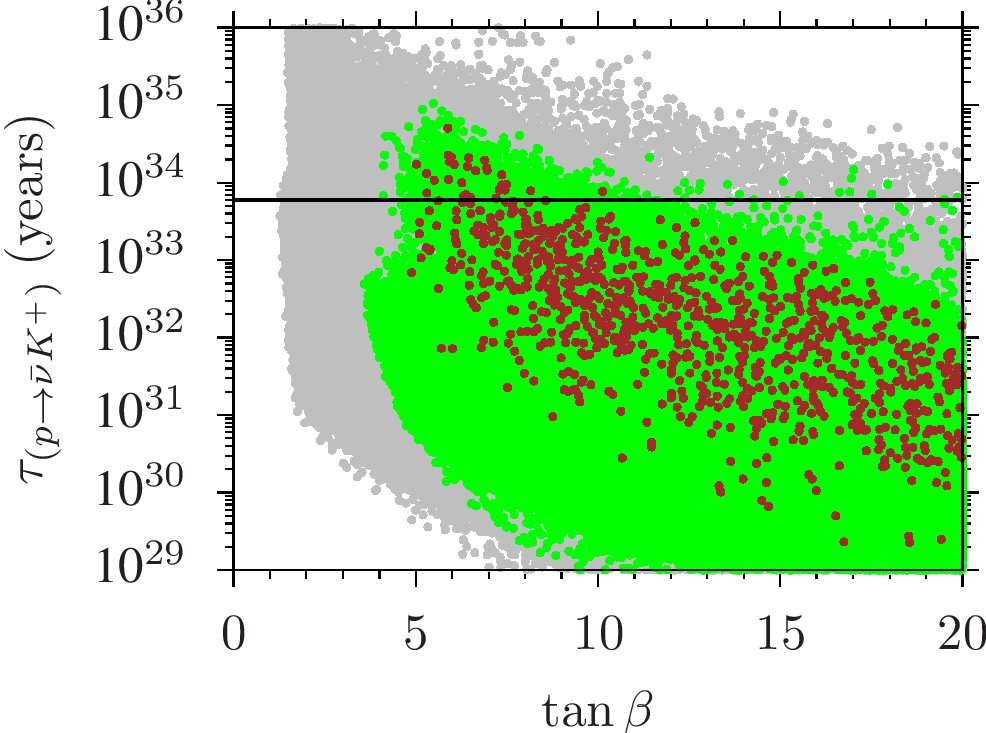}}
\caption{Allowed parameter space for the proton lifetime in  correlation with the fundamental parameters of the model $m_{0_{1,2}}$, $m_{0_{3}}$, $M_{1/2}$ and $\tan\beta$. Proton life time is calculated by setting $M_{H_{C}}=7\times 10^{16}$ GeV. The meaning of colors are the same as given in Figure \ref{fig1}; however, the constraint from the proton decay is not applied in these plots, and the brown points form a subset of green. The horizontal lines indicate the current limit on the proton life time,  $\tau(p\rightarrow \bar{\nu}K^{+})=5.9\times 10^{33}$ years \cite{Abe:2014mwa}. }
\label{fig2}
\end{figure}

\begin{figure}[ht!]
\centering
\subfigure{\includegraphics[scale=1.1]{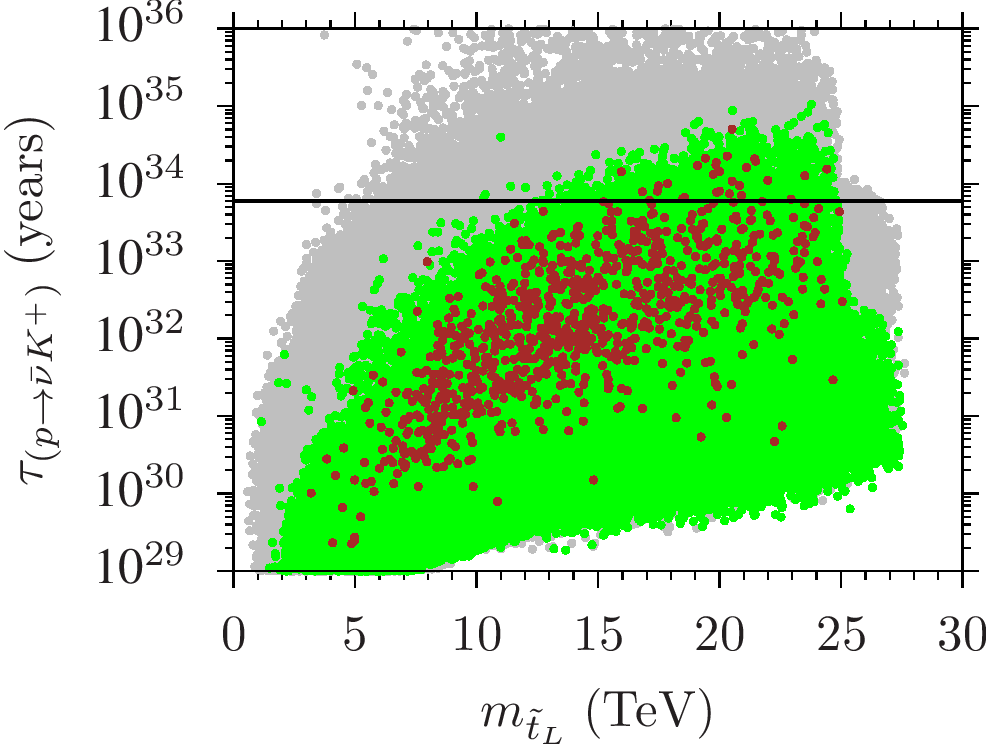}}%
\subfigure{\includegraphics[scale=1.1]{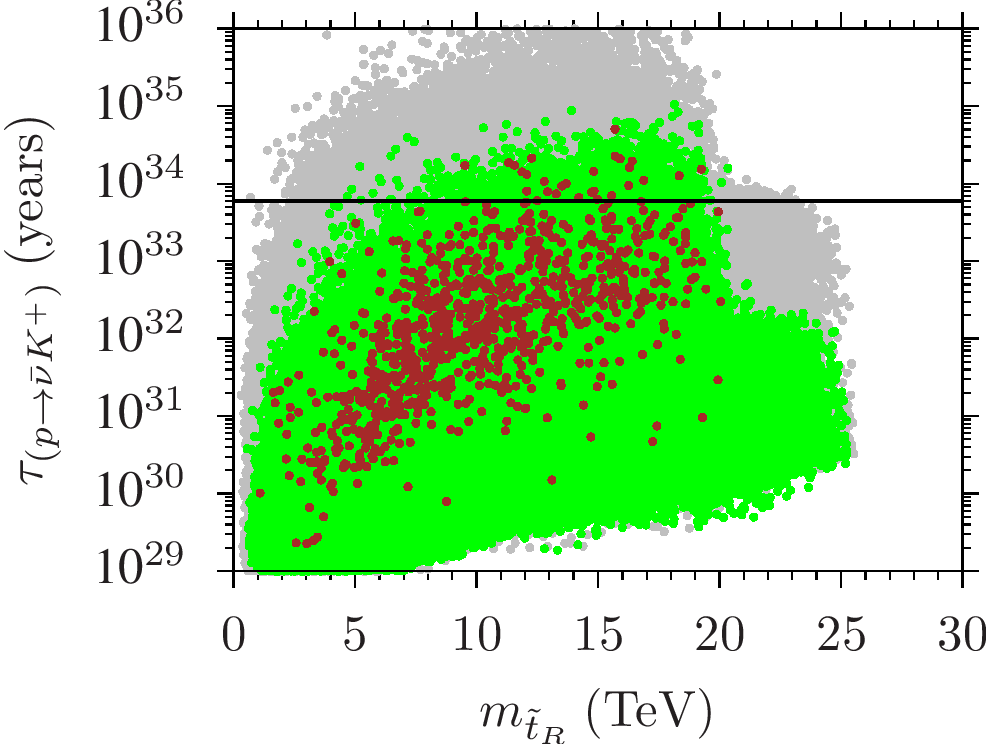}}
\subfigure{\includegraphics[scale=1.1]{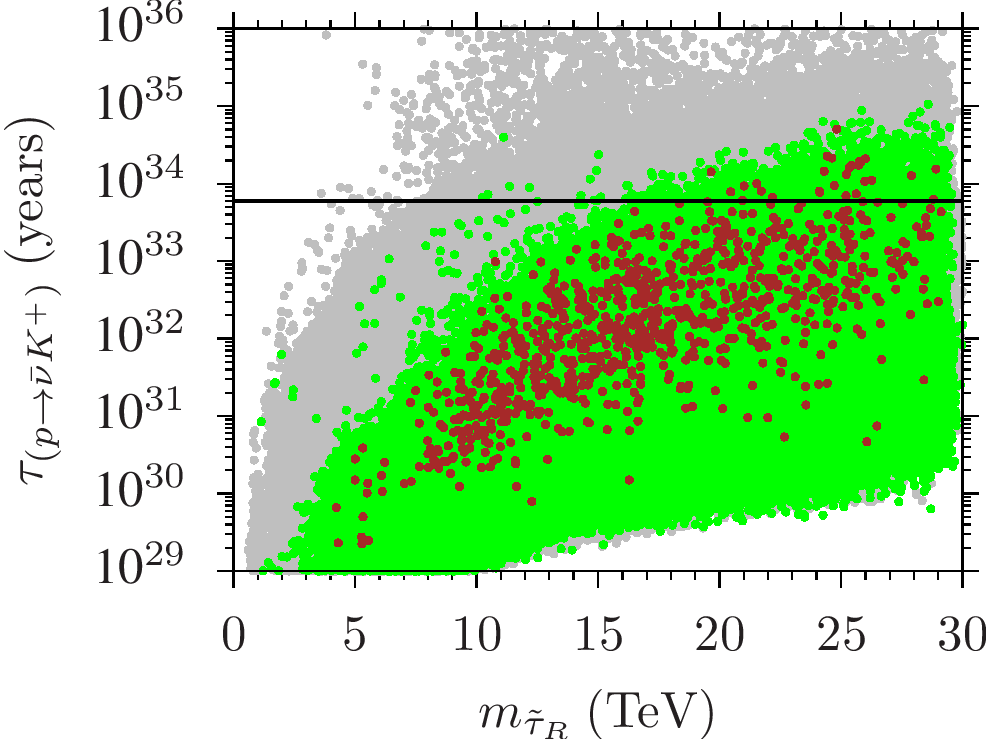}}%
\subfigure{\includegraphics[scale=1.1]{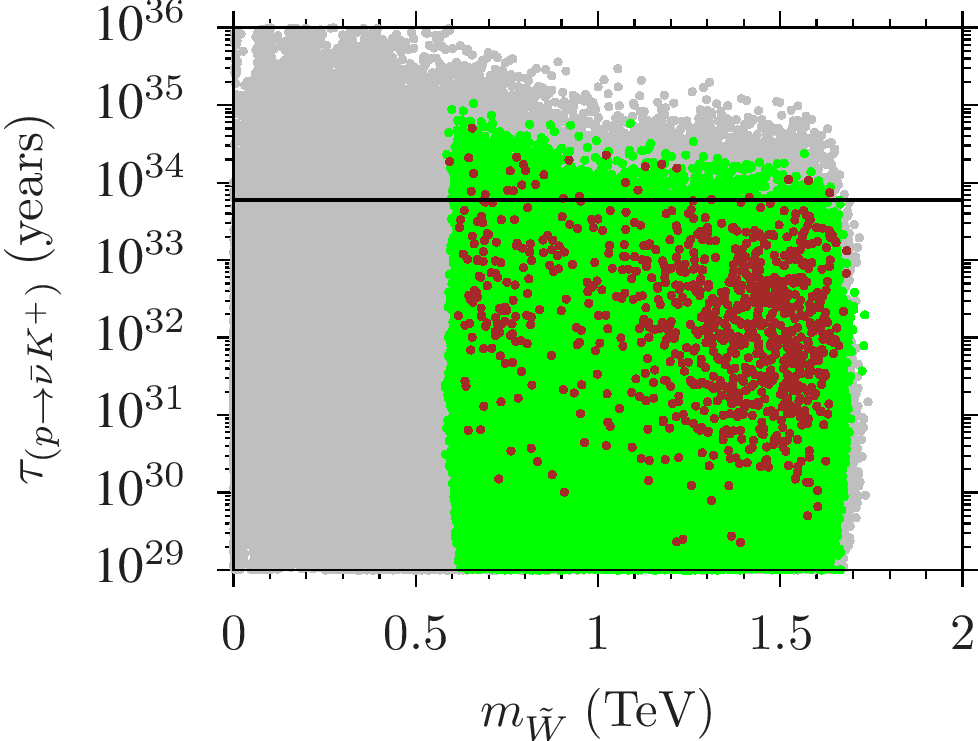}}
\caption{Correlations of proton lifetime with the masses of the left and right-handed stops, right-handed stau and wino respectively. Proton life time is calculated by setting $M_{H_{C}}=7\times 10^{16}$ GeV, and the color coding is the same as Figure \ref{fig2}.}
\label{fig3}
\end{figure}

Fig. \ref{fig2} shows our results for the proton lifetime in correlations with the fundamental parameters of the model, $m_{0_{1,2}}$, $m_{0_{3}}$, $M_{1/2}$ and $\tan\beta$ respectively, assuming that $M_{H_C} = 7 \times 10^{16}$ GeV. The meaning of colors are the same as given in Figure \ref{fig1}; however, the constraint from the proton decay is not applied in these plots, and the brown points form a subset of green. The horizontal line indicates the current limit on the proton life time,  $\tau(p\rightarrow \bar{\nu}K^{+})=5.9\times 10^{33}$ years \cite{Abe:2014mwa}. { The $\tau(p\rightarrow \bar{\nu}K^{+})-m_{0_{1,2}}$ plane shows that many  points  consistent with all collider  constraints (green points) can be  excluded by the current bound on the proton lifetime being below  horizontal line. The longest lifetime in our parameter space can be  $\tau(p\rightarrow \bar{\nu}K^{+})\sim 10^{35}$ years or so.}  The WMAP bound on the relic abundance  bounds the sfermion mass even farther. In contrast to the SSB mass of the first two families, the $\tau(p\rightarrow \bar{\nu}K^{+})-m_{0_{3}}$ plane reveals a strong correlation between the proton lifetime and the SSB mass term for the third family, $m_{0_{3}}$. The solutions consistent with the constraints including that on the proton lifetime requires $m_{0_{3}} \gtrsim 10$ TeV. The reason for such a strong bound is that the third family sparticlse contribute to the proton decay rate proportional to their larger Yukawa couplings. The correlation between the proton lifetime and $M_{1/2}$ is rather weak in the mass range considered, and solutions can be obtained for any value of $M_{1/2}$, once the gluino mass bound is satisfied. Since the proton lifetime is inversely proportional to $\tan\beta$ \cite{Ellis:2015rya}, one can also ameliorate the proton lifetime tension by requiring {small} $\tan\beta$. In the parameter space which we scan over, a strong suppression in proton lifetime is observed  with large $\tan\beta$,  which is expected to be stronger beyond $\tan\beta \geq 20$, disfavoring such large values.

The impact of the proton lifetime on the third family sfermion masses can be seen explicitly from Fig. \ref{fig3} where we present our results for the proton lifetime in correlation with the masses of the left- and right-handed stops, right-handed stau and the wino respectively. The color coding is the same as in Fig. \ref{fig2}. The top panels show that the left-handed stop is mostly required to be heavier than about 7 TeV, while it is also possible to satisfy the constraints with a compatible proton lifetime when $m_{\tilde{t}_{R}} \gtrsim 3$ TeV. The impact becomes stronger for the right-handed stau as is seen from the $\tau(p\rightarrow \bar{\nu}K^{+})-m_{\tilde{\tau}_{R}}$ plane where the solutions with $m_{\tilde{\tau}_{R}} \lesssim 8$ TeV are all excluded by the constraint from the proton lifetime. As is discussed for the SSB gaugino masses, the compatible solutions can be obtained for any $M_{\widetilde{W}}$, once the LHC constraints are satisfied. Thence, if one can suppress the contributions from Higgsino loop, any mass scale for the gauginos can be made consistent down to the value allowed by the current LHC constraints.
\begin{figure}[ht]
\centering
\includegraphics[scale=1.1]{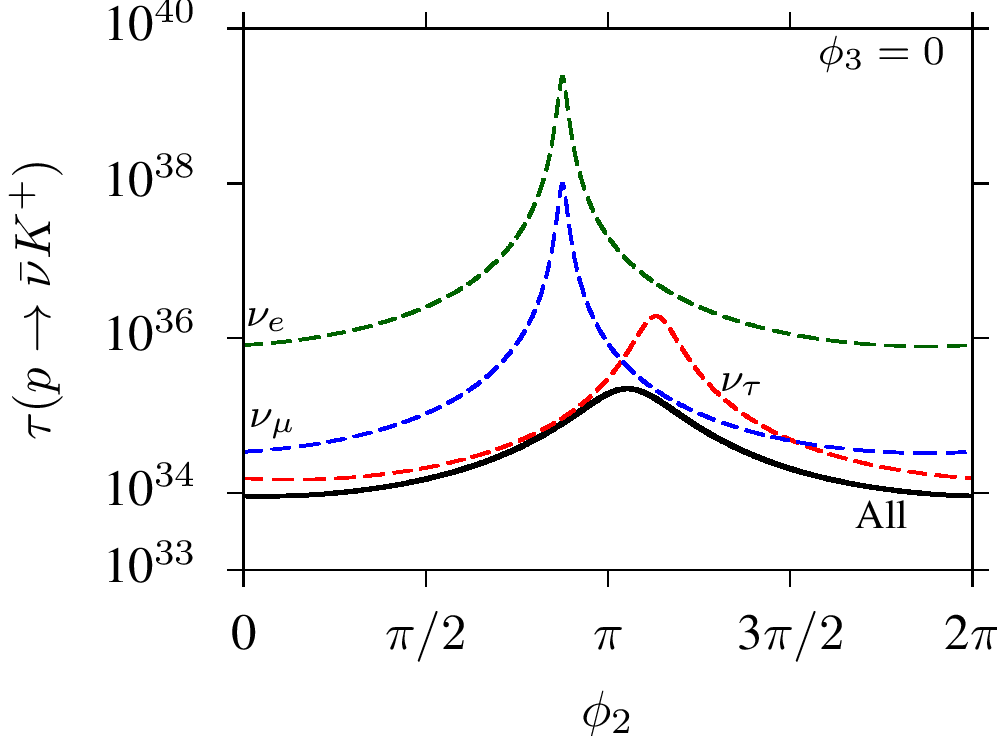}
\caption{Lifetime of the proton in correlation with the phase angle $\phi_2$ with $\phi_3$ set to zero. Green, blue and red dashed curves represent the proton decay into $\nu_{e}$, $\nu_{\mu}$ and $\nu_{\tau}$ along with $K^{+}$, respectively. The solid curve shows the total lifetime of the proton.}
\label{fig:phase}
\end{figure}

Fig. \ref{fig:phase} displays the proton lifetime in correlation with the phase angle, $\phi_{2}$. For simplicity of presentation here we assume $\phi_{3}=0$, which implies $\phi_{1}=-\phi_{2}$, since $\phi_{1}+\phi_{2}+\phi_{3}=0$. Green, blue and red dashed curves represent the proton decay channels  into $\nu_{e}$, $\nu_{\mu}$ and $\nu_{\tau}$ along with $K^{+}$, respectively. The solid  curve shows the total lifetime of the proton. These curves show  that the phase angles can enhance the proton lifetime somewhat. The peaks in $\nu_{e}$ and $\nu_{\mu}$ are observed at $\phi_{2}\simeq 0.87\pi~(\sim 2.7~{\rm in~radian})$, while the peak in $\nu_{\tau}$ is realized at $\phi_{2}\simeq 1.13\pi$ ($\sim 3.55~{\rm in~radian}$). The overall lifetime of the proton peaks at $\phi_{2}\simeq 1.05\pi ~ (\sim 3.31~{\rm in~radian})$.

\begin{figure}[ht!]
\centering
\includegraphics[scale=1.1]{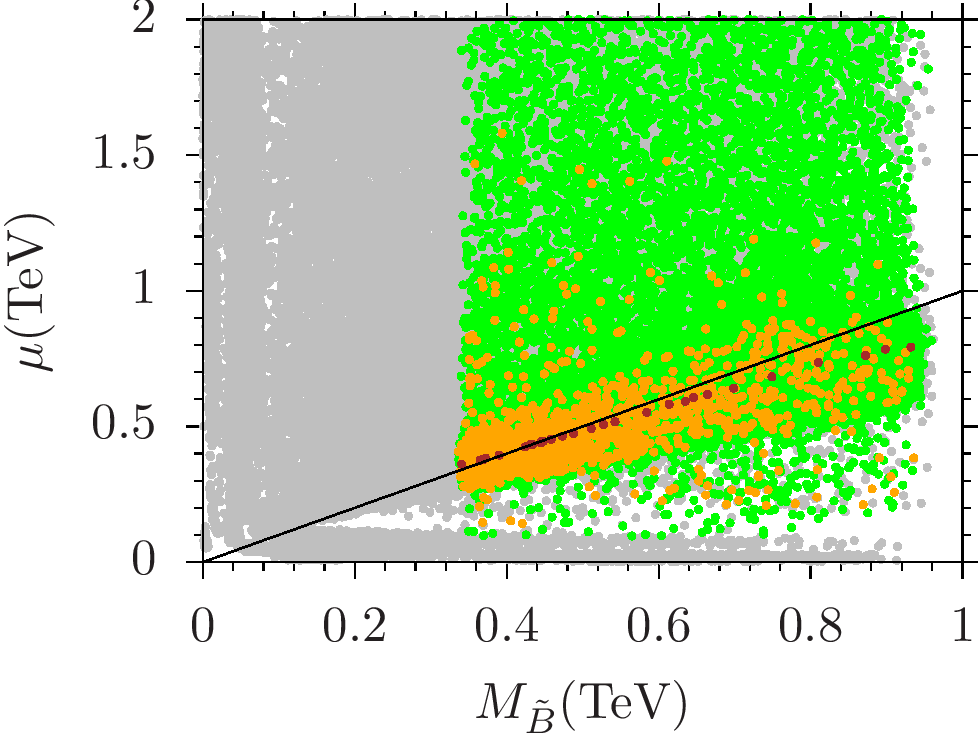}
\caption{Plots for the neutralino species in the $\mu-M_{\tilde{B}}$ plane. All masses plotted here refer to the values at the low scale. The color coding is the same as Figure \ref{fig1}. { The diagonal line shows the solutions in which the Higgsinos and Bino are degenerate in mass ($\mu = M_{\tilde{B}}$).}}
\label{fig4}
\end{figure}

\begin{figure}[ht]
\subfigure{\includegraphics[scale=1.1]{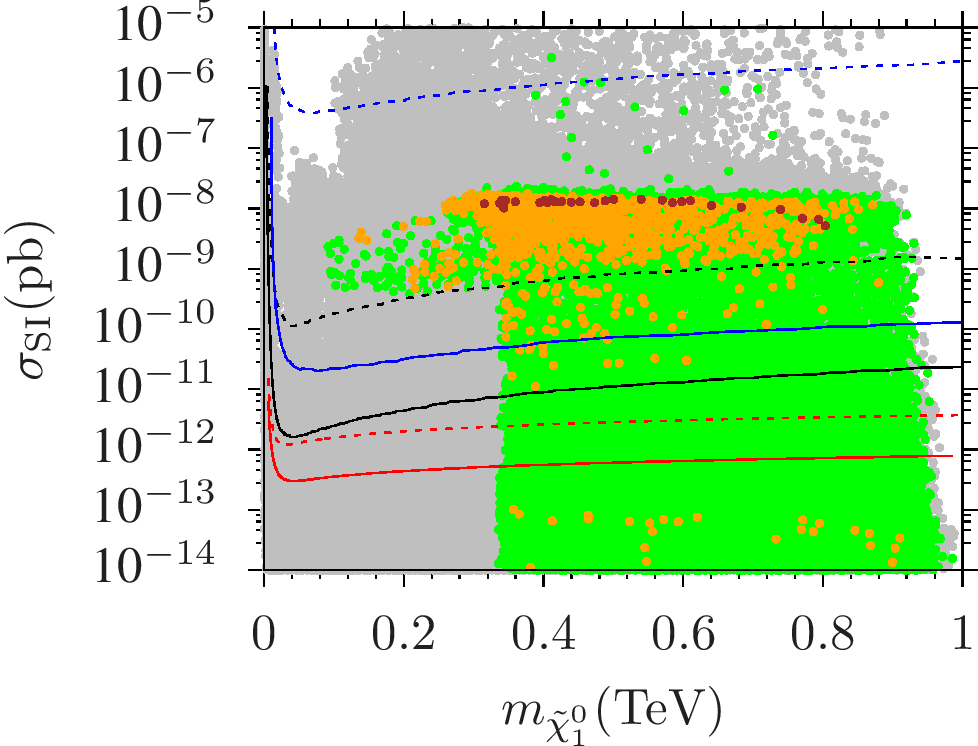}}
\subfigure{\includegraphics[scale=1.1]{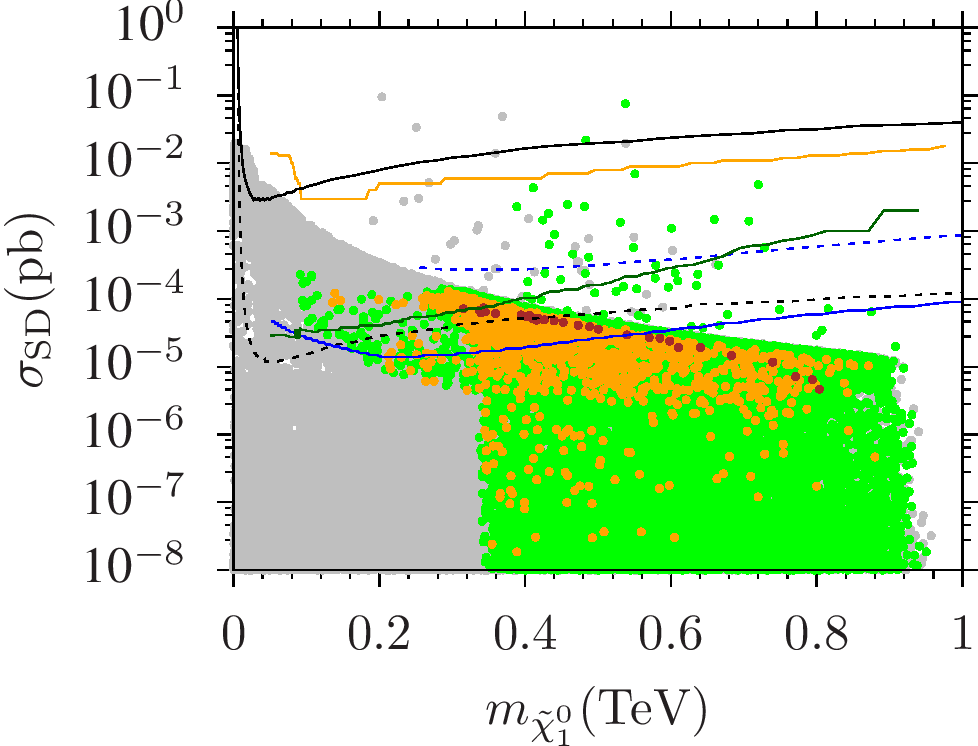}}
\caption{Plots for the spin-independent (left) and spin-dependent (right) scattering cross-sections of the DM scattering off nuclei. The color coding is the same as Figure \ref{fig1}. In the $\sigma_{SI}-m_{\tilde{\chi}_{1}^{0}}$ plane, the blue dashed (solid) line represents the current (future) exclusion from the CDMS experiment \cite{Brink:2005ej,Aramaki:2016spe}, while the black dashed (solid) line indicates the current (projected) results from the LUX (LZ) experiment \cite{Akerib:2018lyp}. The red dashed (solid) line displays the current (future) exclusion curve from the XENON1T (XENONnT) experiment \cite{Aprile:2020vtw}. In the $\sigma_{SD}-m_{\tilde{\chi}_{1}^{0}}$ plane, the black solid line represents the current results from the LUX experiment \cite{Akerib:2016lao}, while the orange solid line indicates the current exclusion from the SuperK measurements \cite{Tanaka:2011uf}. The blue dashed (solid) line stands for the current (future) sensitivity of the IceCube experiment \cite{IceCube:2009iyf}. Finally the green solid line is provided by the CMS experiment at 8 TeV \cite{Khachatryan:2014rra}. }
\label{fig5}
\end{figure}

We present our results for the masses of the neutralino species in Fig. \ref{fig4} with a plot in the $\mu-M_{\tilde{B}}$ plane. All masses plotted refer at their low scale values. The color coding is the same as Fig. \ref{fig1}. { The diagonal line shows the solutions in which the Higgsinos and Bino are degenerate in mass ($\mu = M_{\tilde{B}}$).} Since it is possible to realize $\mu-$term below about 1 TeV, the Higgsinos can be the significant component of the DM, and as is seen from the $\mu-M_{1}$ plane, the Higgsinos can either be degenerate with Bino or lighter than it. Such solutions yield either Bino-Higgsino mixture in the DM formation, or mostly Higgsino DM. { The diagonal line guides us to see realization of  bin- Higgsino dark matter in our parameter space.}

When the DM composition involves a significant amount of Higgsinos, it yields large cross-sections for the DM scattering off nuclei, since these processes happen through Yukawa interactions. In this context, the DM predictions of our model {receive a strong constraint from} the direct DM detection experiments as plotted in Fig. \ref{fig5} for the spin-independent (left) and spin-dependent (right) scattering cross-sections. The color coding is the same as Fig. \ref{fig1}. In the $\sigma_{SI}-m_{\tilde{\chi}_{1}^{0}}$ plane, the blue dashed (solid) line represents the current (future) exclusion from the CDMS experiment \cite{Brink:2005ej,Aramaki:2016spe}, while the black dashed (solid) line indicates the current (projected) results from the LUX (LZ) experiment \cite{Akerib:2018lyp}. The red dashed (solid) line displays the current (future) exclusion curve from the XENON1T (XENONnT) experiment \cite{Aprile:2020vtw}. In the $\sigma_{SD}-m_{\tilde{\chi}_{1}^{0}}$ plane, the black solid line represents the current results from the LUX experiment \cite{Akerib:2016lao}, while the orange solid line indicates the current exclusion from the SuperK measurements \cite{Tanaka:2011uf}. The blue dashed (solid) line stands for the current (future) sensitivity of the IceCube experiment \cite{IceCube:2009iyf}. Finally the green solid line is provided by the CMS experiment at 8 TeV \cite{Khachatryan:2014rra}. The $\sigma_{SI}-m_{\tilde{\chi}_{1}^{0}}$ plane shows that most of the solutions yield large spin-independent cross-sections so that they are slightly above the exclusion limit from the current LUX experiment, while the XENON experiment reveals a stronger impact on the results, since its sensitivity has recently been significantly improved. On the other hand, these solutions are in the reach of the projected results from the SuperCDMS experiment \cite{Aramaki:2016spe}, and they are expected to be excluded or discovered in near future.

The dark matter searches provide strong constraints on the parameters of the model. Even though the experiments provide model independent results, the phenomenological analyses are rather model dependent and based on strict assumptions. In our analyses we have assumed the dark matter relic density is saturated only by the LSP neutralino. {With this assumption, the model under consideration predicts large scattering cross-sections for the dark matter scattering off nuclei, which are excluded by several direct detection experiments such as LUX (black dashed curve) and XENON1T (red dashed curve).} Thus the assumption about dark matter composition needs some modification.  We note that it is easy to satisfy the upper limit on LSP abundance from over-closing the universe.  If the LSP contributes only a fraction of the DM abundance, there is no issue with the model.  This can be realized, for example, if the axion contributes the remainder of the DM abundance.  Inclusion of the axion and its SUSY partners do not significantly modify the phenomenology discussed here.


\begin{table}[ht!]
\centering
\setstretch{1.2}
\begin{tabular}{|c|ccc|}
\hline  & Point 1 & Point 2 & Point 3 \\ \hline
$m_{0_{1,2}}$ & {12580} & 16720 & 21050 \\
$m_{0_{3}}$ & 25750 & 26140 & 26110 \\
$M_{1/2}$ & 815.3 & 803 & 1947 \\
$A_{0}/m_{0_{3}}$ & -0.746 & 1.4 & -1.12 \\
$\mu$ & 394.8 & 360 & 792.1 \\
$m_{A}$ & 14730 & 26170 & 20420 \\
$m_{t}$ & 173.3 & 173.3 & 173.3 \\
$\tan\beta$ & 6.38 & 5.93 & 5.31 \\ \hline
$m_{h}$ & 124.19 & 125.35 & 123.22 \\
$m_{H}$ & 14827 & 26342 & 20554 \\
$m_{A}$ & 14730 & 26170 & 20420 \\
$m_{H^{\pm}}$ & 14730 & 26170 & 20420 \\ \hline
$m_{\tilde{\chi}_{1}^{0}}$, $m_{\tilde{\chi}_{2}^{0}}$ & 359.58, 412.35 & 315.49, 375.83 & 803.66, 817.58 \\
$m_{\tilde{\chi}_{3}^{0}}$, $m_{\tilde{\chi}_{4}^{0}}$ & 425.25, 769.71 & 382.82, 671.81 & 937.27, 1781 \\
$m_{\tilde{\chi}_{1}^{\pm}}$, $m_{\tilde{\chi}_{2}^{\pm}}$ & 418.8, 741.16 & 377.34, 647.93 & 845.5, 1731 \\
$m_{\tilde{g}}$ & 2275 & 2102 & 4848 \\ \hline
$m_{\tilde{u}_{L,R}}$, & 12216, 13068 & 16513, 16845 & 21103, 21551 \\
$m_{\tilde{t}_{1,2}}$, & 15255, 21197 & 10955, 20021 & 12962, 20630 \\  \hline
$m_{\tilde{d}_{L,R}}$, & 12217, 12158 & 16514, 16567 & 21103, 21091 \\
$m_{\tilde{b}_{1,2}}$, & 21136, 25727 & 20107, 26129 & 20659, 26164 \\  \hline
$m_{\tilde{\nu}_{1,2}}$ & 12953 & 16814 & 21292 \\
$m_{\tilde{\nu}_{3}}$ & 25954 & 26202 & 26331 \\
$m_{\tilde{e}_{L,R}}$, & 12974, 11610 & 16809, 16403 & 21282, 20569 \\
$m_{\tilde{\tau}_{1,2}}$, & 25224, 25954 & 25817, 26171 & 25644, 26277 \\ \hline
$\sigma_{SI}$ (pb) & $ 1.28 \times 10^{-8} $ & $ 1.18 \times 10^{-8} $ & $ 5.13 \times 10^{-9} $ \\
$\sigma_{SD}$ (pb) & $ 6.11 \times 10^{-5} $ & $ 7.12 \times 10^{-5} $ & $ 4.65 \times 10^{-6} $ \\
$\Omega h^{2}$ & {0.122} & {0.12} & {0.117} \\ \hline
$\tau(p\rightarrow \bar{\nu}K^{+})\times 10^{-33}$ & 6.99 & 18.56 & 7.36 \\ \hline
\end{tabular}
\caption{ A table of three benchmark points satisfying the mass bounds, B-physics constraints and the Proton lifetime measurements. All the masses are in units of GeV, and the proton lifetime is given in years. Points 1 displays the lowest value for $m_{0_{1,2}}$ when solutions are allowed by the Planck bound within $5\sigma$. Point 2 shows a solution for the lightest Higgsino mass consistent with the Planck bound. Point 3 represents solutions with relatively lower spin-independent and spin-dependent scattering cross-sections of DM.}
\label{tab1}
\end{table}

{Finally we present a table of three benchmark points in Table \ref{tab1}, which exemplify our findings. All points are chosen as to be consistent with the mass bounds, $B$-physics constraints and the proton lifetime measurements. If one requires the solutions to be consistent with the Planck bound on the relic density of LSP, then the minimum value for the SSB scalar masses of the first two-family matter fields are observed to be $m_{0_{1,2}} \simeq 12.6$ TeV  as exemplified with Point 1. In addition, Point 2 displays a solution for the lightest Higgsino compatible with the Planck bound on relic density of LSP neutralino, and Point 3 represents solutions with relatively lower spin-independent and spin-dependent scattering cross-sections of DM. In addition to the light Higgsinos revealed in all the benchmark points, Bino also happens to be as light as about the Higgsino, and it results in Bino-Higgsino mixture in DM composition. {If the DM is composed by Higgsinos or it happens to be Bino-Higgsino mixture, then the solutions typically lead to chargino-neutralino coannihilation scenarios}.}

\section{\boldmath{$d=6$} Proton  Decay}
\label{sec:d=6}
Here we consider the proton decay rate induced by the exchange of the $SU(5)$ gauge $X,Y$ bosons (d=6 proton decay). The dominant decay channel in this case is $p \rightarrow e^+ \pi^0$.

The effective K$\ddot{a}$hler potential for dimension-six operators is given by
\begin{eqnarray}
{\cal L}_{\text{dim.6}} = & \int d^4 \theta \left( \sum_{i=1}^2 C^{(i)} {\cal O}^{(i)} + {\rm h.c.} \right)\label{eq:dim_6},
\end{eqnarray}
with operators ${\cal O}^{(i)}~(i=1,2)$ defined as
\begin{eqnarray}
{\cal O}^{(1)} & = \epsilon_{\alpha\beta\gamma} \epsilon_{rs} U^{C \dag \alpha} D^{C\dag \beta}  Q^{r \, \gamma} L^s , \,\,\,
{\cal O}^{(2)} = \epsilon_{\alpha\beta\gamma} \epsilon_{rs} E^{C\dag} U^{C \, \dag \, \alpha} Q^{r\beta} Q^{s\gamma}.
\end{eqnarray}
Here for simplicity  we omitted  the flavor indices. The Wilson coefficients $C_{\rm GUT}^{(i)}$ are defined as
\begin{eqnarray}
C_{\rm GUT}^{(1)} & = C_{\rm GUT}^{(2)} = - \frac{g_5^2}{M_X^2}.
\end{eqnarray}
%
Note that the Wilson coefficients  at low energies do not depend explicitly on the masses of SUSY  particles, in contrast to those of the dimension-five proton decay operators.

The partial decay width for $p \to e^+ \pi^0$ is then given by \cite{Ellis:2019fwf}:
\begin{equation}
 \Gamma (p\to  \pi^0 e^+)=
\frac{m_p}{32\pi}\biggl(1-\frac{m_\pi^2}{m_p^2}\biggr)^2
\bigl[
\vert {\cal A}_L(p\to \pi^0 e^+) \vert^2+
\vert {\cal A}_R(p\to \pi^0 e^+) \vert^2
\bigr]~,
\end{equation}
with
\begin{align}
 {\cal A}_L(p\to \pi^0 e^+)&=
- \frac{g_5^2}{M_X^2} \cdot
A_1 \cdot \langle \pi^0\vert (ud)_Ru_L\vert p\rangle
~,\nonumber \\
 {\cal A}_R(p\to \pi^0 e^+)&=
- \frac{g_5^2}{M_X^2} (1+|V_{ud}|^2) \cdot
A_2 \cdot \langle \pi^0\vert (ud)_Ru_L\vert p\rangle
~,
\end{align}
where $A_{1}\simeq 2.72$ and $A_{2}\simeq 3.08$ are the renormalization factors \cite{Ellis:2019fwf}. As mentioned in Section \ref{sec3} from low energy data with RGE extrapolation we can determine the effective mass $M_G = (M^2_XM_{\Sigma})^{1/3}$.  The full range of this mass parameter is given in Fig. \ref{fig-RGE444}. To a
good approximation we can wrie down the $d=6$ proton decay inverse rate as
\begin{equation}
    \tau (p \to e^+ \pi^0) \simeq 1.8 \times 10^{35}~{\rm yr.}
    \times \biggl(\frac{M_X}{10^{16}~{\rm GeV}}\biggr)^4 ~.
\end{equation}


\section{Conclusion}
\label{sec:conc}

We have presented in this paper a re-appraisal of the proton lifetime in minimal SUSY $SU(5)$ grand unified theory.  The particle content of the model is kept minimal, with three families of ${\bf 10} + {\bf \overline{5}}$ fermions and a Higgs sector consisting of a {\bf 24} and a pair of ${\bf 5 + \overline{5}}$. We have incorporated realistic fermion masses by including Planck-suppressed $d=5$ operator in the Yukawa coupling sector.  This leads to a decrease in the proton lifetime rate by a factors of about 20 and thus constrains the SUSY parameter space even more.  We have also included Planck-suppressed operators that smear the unified gauge coupling of $SU(5)$.  These operators, along with $d=5$ operators arising from the symmetry breaking sector, are shown to help raise the mass of the color-triplet Higgsino to about $7 \times 10^{16}$ GeV.  This counterbalances somewhat the enhanced proton decay rate resulting from realistic fermion masses.

We have also paid close attention to the SUSY parameter space. Our framework allows for a universal mass for the first two family sfermions that is different from that of the third family sfermons.  Such a spectrum is motivated by flavor symmetry based MSSM \cite{Babu:2014sga,Babu:2014lwa}.  We have allowed the scalar masses to be as large as 30 TeV, so that the direct search limits from the LHC can be satisfied, along with constraints arising from proton lifetime.  The gaugino mass parameter is however limited to $M_{1/2} < 2$ TeV, so that there is a consistent dark matter candidate.  Such a spectrum {opens} the possibility that the gauginos and the Higgsinos may be within reach of the high luminosity run of the LHC.  We have also elucidated expectations for dark matter searches through its spin-dependent and spin-independent scattering off nucleons. When all the constraints of the model are imposed we find that the lifetime for proton decaying into $\overline{\nu} + K^+$ is likely to be shorter than about $10^{35}$ yrs.

\section{Acknowledgments}

We thank A. Ismail, N. Nagata and K. A. Olive  for helpful discussions. {We also thank an anonymous referee for pointing out an error in the numerical results in an earlier version of the draft.} The work of KSB is supported in part by the US Department of Energy Grant No. DE-SC 0016013. The research of C.S.U. was supported in part by the Spanish MICINN, under grant PID2019-107844GB-C22.

\appendix

\section{$d=5$ Proton Decay calculation }

In this Appendix we provide the steps followed for computing the $d=5$ proton decay rate within our framework.  We  have followed closely the steps outlined in Ref. \cite{Ellis:2015rya}. The only difference in our approach is that we do not use bottom-tau Yukawa coupling unification condition at GUT scale  -- as this condition is modified by Planck-induced threshold corrections in our framework.  As shown in Eq. (\ref{massnew}) in our scenario the down quark and charge lepton  Yukawa couplings are independent of each other.

The effective Lagrangian obtained after integrating out the color-triplet Higgsino fields is written as
\begin{equation}
\mathcal{L}^{{\rm eff}}_{5} = C_{5L}^{ijkl}\mathcal{O}^{5L}_{ijkl}+C_{5R}^{ijkl}\mathcal{O}^{5R}_{ijkl}+{\rm h.c.}
\end{equation}
with
\begin{equation}
\begin{array}{l}
\mathcal{O}^{5L}_{ijkl} = \int d^{2}\theta\dfrac{1}{2}\epsilon_{abc}(Q^{a}_{i}\cdot Q^{b}_{j})(Q^{c}_{k}\cdot L_{l}) \\ \\
\mathcal{O}^{5L}_{ijkl} = \int d^{2}\theta\epsilon^{abc}\bar{u}_{ia}\bar{e}_{j}\bar{u}_{kb}\bar{d}_{lc}~.
\end{array}
\end{equation}
The Wilson coefficients are given at the GUT scale as
\begin{equation}
\begin{array}{l}
C_{5L}^{ijkl}(\mgut) = \dfrac{1}{M_{H_{C}}}h_{\mathbf{10},i}\delta^{ij}V^{*}_{kl}h_{\mathbf{\bar{5}},l}, \\ \\
C_{5R}^{ijkl}(\mgut) =\dfrac{1}{M_{H_{c}}}h_{\mathbf{10},i}V_{ij}V^{*}_{kl}h_{\mathbf{\bar{5}},l}^{\prime},
\end{array}
\label{eq:wilsons}
\end{equation}
where $V_{ij}$ are the CKM matrix elements parametrized as
\begin{equation}\hspace{-0.8cm}
V_{{\rm CKM}} = \left( \begin{array}{ccc}
1-\lambda^{2}/2 & \lambda & A\lambda^{3}(\rho - i\eta) \\
-\lambda & 1-\lambda^{2}/2 & A\lambda^{2} \\
A\lambda^{3}(1-\rho -i\eta) & -A\lambda^{2} & 1
\end{array}\right) \hspace{0.5cm}{\rm with}\hspace{0.5cm} \left\lbrace \begin{array}{l}
A = 0.818 \\
\lambda=0.2255 \\
\rho=0.117 \\
\eta=0.353
\end{array}\right.
\end{equation}

For the Yukawa couplings, we use at the GUT scale the tree-level matching conditions. However, we note here that there is an ambiguity in the determination of the GUT Yukawa couplings. As is known, the $b-\tau$ Yukawa unification in the SUSY $SU(5)$ is not a good fit in most of the parameter space \cite{Gogoladze:2011be}. The inclusion of higher dimensional operators cures this problem in our framework, see Eq. (\ref{massnew}). As a result we have the following GUT scale matching condition for Yukawa couplings:
\begin{equation}
h_{\mathbf{10}i}=\dfrac{1}{4}f_{u_i},\hspace{0.5cm} h_{\mathbf{\bar{5}}i}=\sqrt{2} f_{e_{i}},\hspace{0.5cm} h_{\mathbf{\bar{5}}i}^{\prime}=\sqrt{2} f_{d_{i}}.
\end{equation}
Here $i=1,2,3$ is the family index. {For the third generation we use top, bottom and tau Yukawa couplings obtained through ISAJET RGE running, which are approximately $f_{t}=89.1/v_{u}$, $f_{b}=0.96/v_{d}$, and $f_{\tau}=1.33/v_{d}$ for most values of $\tan\beta$ we investigate (with $v_{u,d}$ in GeV). For the first two generation quark and lepton Yukawa couplings  we use their  GUT scale value obtained in Ref. \cite{Babu:2018tfi}:
$f_c=0.245/v_{u}$, $f_s=18.46\times 10^{-3}/v_{d}$, $f_{\mu}=80.04\times 10^{-3}/v_{d}$, $f_u=0.45\times 10^{-3}/v_{u}$, $f_d=0.585\times 10^{-3}/v_{d}$, $f_e=0.379\times 10^{-3}/v_{d}$; with $v_{u,d}$ in GeV. Note that $v_{d}=v_{{\rm SM}}\cos\beta$ and $v_{u}=v_{{\rm SM}}\sin\beta$ with $v_{{\rm SM}} = 174$ GeV.}

Following the discussions of Sec. \ref{sec3}. we fix the color-triplet Higgsino mass at
\begin{equation}
M_{H_{c}}=7\times 10^{16}\, GeV~.
\end{equation}
Such a value is compatible with the model. For any other values one can simply rescale the results on proton lifetime.

In our calculations we parameterize the Yukawa couplings as follows:
\begin{equation}
(h_{\mathbf{10}})_{ij} = e^{i\phi_{i}}\delta_{ij}h_{\mathbf{10},i},\hspace{0.3cm}(h_{\mathbf{\bar{5}}})_{ij} = V^{*}_{ij}h_{\mathbf{\bar{5}},i}
\end{equation}
where $\phi_{i}$ are the unknown $SU(5)$ phases obeying the condition $\phi_{1}+\phi_{2}+\phi_{3}=0$. For most of our calculations we set $\phi_{i}=0$ for simplicity, although we have studied the dependence of proton lifetime on one of the phases as shown in Fig. \ref{fig:phase}.

 At the scale of SUSY breaking, $M_{{\rm SUSY}}$, the sfermions in the dimension-5 operators are integrated out by evaluating the loop diagrams involving the Higgginos and Wino.  The
dominant baryon number violating interactions after this integration are given by \cite{Ellis:1983qm,Belyaev:1982ik}
\begin{eqnarray}
\mathcal{L}^{{\rm eff}}_{6} = C_{i}^{\tilde{H}(3)}\mathcal{O}_{1i33}+C_{i}^{\tilde{H}(2)}\mathcal{O}_{1i22}+C_{jk}^{\tilde{W}}\tilde{\mathcal{O}}_{1jjk}+C_{jk}^{\tilde{W}}\tilde{\mathcal{O}}_{j2jk}+\overline{C}_{jk}^{\tilde{W}}\tilde{\mathcal{O}}_{jj1k}~.
\label{eq:Leff6}
\end{eqnarray}
with
\begin{equation}
\setstretch{1.5}
\begin{array}{ll}
\mathcal{O}_{ijkl} & = \epsilon_{abc}(u_{Ri}^{a}d_{Rj}^{b})(Q_{Lk}^{c}L_{Ll})~, \\
\tilde{\mathcal{O}}_{ijkl} & = \epsilon_{abc}\varepsilon^{\alpha\beta}\varepsilon^{\gamma\delta}(Q^{a}_{Li\alpha}Q^{b}_{Lj\gamma})(Q^{c}_{Lk\delta}L_{Ll\beta})~.
\end{array}
\end{equation}

The first two terms in Eq. (\ref{eq:Leff6}) are the Higgsino contributions, while the other terms represent the wino contribution to the proton decay rate.
%
%
Note that the Wilson coefficients given in Eq. (\ref{eq:wilsons}) are calculated with $C_{5L}$ and $C_{5R}$ at $M_{{\rm SUSY}}$. Once they are obtained at the GUT scale, their values at $M_{{\rm SUSY}}$ can be obtained through the renormalization group equations given by \cite{Ellis:2015rya}:
\begin{eqnarray}
&&\frac{d}{d\ln Q} C^{ijkl}_{5L} =
\frac{1}{16\pi^2}\left[-\frac{2}{5}g_1^2 -6g_2^2 -8g_3^2
+y_{u_i}^2 +y_{d_i}^2 + y_{u_j}^2 +y_{d_j}^2 +
y_{u_k}^2 + y_{d_k}^2 + y_{e_l}^2
\right] C^{ijkl}_{5L}, \nonumber \\
 &&\frac{d}{d\ln Q} C^{ijkl}_{5R} =
\frac{1}{16\pi^2}\left[-\frac{12}{5}g_1^2 -8g_3^2
+2y_{u_i}^2 + 2y_{e_j}^2 + 2y_{u_k}^2 + 2y_{d_l}^2
\right] C^{ijkl}_{5R}.
\end{eqnarray}
Here $Q$ denotes the renormalization scale. The Matching conditions at the SUSY scale are

\begin{equation}
\setstretch{2.5}
\begin{array}{ll}
C_i^{\widetilde{H}(3)}(M_{SUSY}) & = \frac{f_t f_{\tau}}{(4\pi)^2}C^{*331i}_{5R}(M_{SUSY})F(\mu, m^2_{\tilde{t}_R}m^2_{\tilde{\tau}_R}) \\
C_i^{\widetilde{H}(2)}(M_{SUSY}) & = \frac{f_c f_{\mu}}{(4\pi)^2}C^{*221i}_{5R}(M_{SUSY})F(\mu, m^2_{\tilde{c}_R}m^2_{\tilde{\mu}_R})
\end{array}
\end{equation}
where the Yukawa couplings $f_t$, $f_{\tau}$, $f_{c}$ and $f_{\mu}$ take their SUSY scale values. In addition, we have
\begin{eqnarray}
C_{jk}^{\widetilde{W}}(M_{SUSY})=\frac{\alpha_2}{(4\pi)} C^{jj1k}_{5L}(M_{SUSY})[F(M_2, m^2_{\tilde{Q}_1}m^2_{\tilde{Q}_j}+ F(M_2, m^2_{\tilde{Q}_j}m^2_{\tilde{L}_k})],
\end{eqnarray}
\begin{eqnarray}
\overline{C}_{jk}^{\widetilde{W}}(M_{SUSY})=\frac{3}{2}\frac{\alpha_2}{(4\pi)} C^{jj1k}_{5L}(M_{SUSY})[F(M_2, m^2_{\tilde{Q}_j}m^2_{\tilde{Q}_j})+ F(M_2, m^2_{\tilde{Q}_1}m^2_{\tilde{L}_k})],
\end{eqnarray}
where we used the standard notation for MSSM sfermions and $\alpha_i\equiv g_i^2/4\pi$.
The  loop function $F(M,m_{1}^{2},m_{2}^{2})$ is given by
\begin{eqnarray}
F(M, m^2_1 m^2_2) \equiv \frac{M}{m_1^2 - m_2^2}\left[\frac{m_1^2}{m_1^2-M^2}ln\left(\frac{m_1^2}{M^2}\right)- \frac{m_2^2}{m_2^2-M^2}ln\left(\frac{m_2^2}{M^2}\right)\right]
\end{eqnarray}

{The values of the Wilson coefficients at the electroweak scale can be obtained through the following RGEs \cite{Alonso:2014zka}}:
 \begin{align}
 \frac{d}{d\ln Q} C^{\widetilde{H}}_{i}
&= \biggl[\frac{\alpha_1}{4\pi}\biggl(-\frac{11}{10}\biggr)
+\frac{\alpha_2}{4\pi}\biggl(-\frac{9}{2}\biggr)
+\frac{\alpha_3}{4\pi}(-4) +\frac{1}{2}\frac{f_{t}^2}{16\pi^2}
\biggr]C^{\widetilde{H}}_{i} ,\nonumber \\
 \frac{d}{d\ln Q} C^{\widetilde{W}}_{jk}
&= \biggl[\frac{\alpha_1}{4\pi}\biggl(-\frac{1}{5}\biggr)
+\frac{\alpha_2}{4\pi}(-3)
+\frac{\alpha_3}{4\pi}(-4) +\frac{f_{u_j}^2}{16\pi^2}
\biggr]C^{\widetilde{W}}_{jk}
+\frac{\alpha_2}{4\pi}(-4)[
2C^{\widetilde{W}}_{jk} + \overline{C}^{\widetilde{W}}_{jk}] ,\nonumber \\
 \frac{d}{d\ln Q} \overline{C}^{\widetilde{W}}_{jk}
&= \biggl[\frac{\alpha_1}{4\pi}\biggl(-\frac{1}{5}\biggr)
+\frac{\alpha_2}{4\pi}(-3)
+\frac{\alpha_3}{4\pi}(-4) +\frac{f_{u_j}^2}{16\pi^2}
\biggr]\overline{C}^{\widetilde{W}}_{jk}
+\frac{\alpha_2}{4\pi}(-4)[
2C^{\widetilde{W}}_{jk} + \overline{C}^{\widetilde{W}}_{jk}] ,
\end{align}
{where $f_{u_j}$ denote the SM up-type Yukawa couplings. The effective operators inducing the $p\to K^+\bar{\nu}_k$ decay mode, and corresponding interactions can be written as}

 \begin{align}
 {\cal L}(p\to K^+\bar{\nu}_i^{})
=&C_{RL}(usd\nu_i)\bigl[\epsilon_{abc}(u_R^as_R^b)(d_L^c\nu_i^{})\bigr]
+C_{RL}(uds\nu_i)\bigl[\epsilon_{abc}(u_R^ad_R^b)(s_L^c\nu_i^{})\bigr]
\nonumber \\
+&C_{LL}(usd\nu_i)\bigl[\epsilon_{abc}(u_L^as_L^b)(d_L^c\nu_i^{})\bigr]
+C_{LL}(uds\nu_i)\bigl[\epsilon_{abc}(u_L^ad_L^b)(s_L^c\nu_i^{})\bigr] ,
\end{align}
{where,}
\begin{align}
 C_{RL}(usd\nu_\tau)&=-V_{td}C^{\widetilde{H}(3)}_{2}(m_Z),\nonumber \\
 C_{RL}(uds\nu_\tau)&=-V_{ts}C^{\widetilde{H}(3)}_{1}(m_Z),\nonumber \\
C_{RL}(usd\nu_\mu)&=-V_{cd}C^{\widetilde{H}(2)}_{2}(m_Z),\nonumber \\
 C_{LL}(usd\nu_k)&=\sum_{j=2,3}V_{j1}V_{j2}
C^{\widetilde{W}}_{jk}(m_Z),\nonumber \\
 C_{LL}(uds\nu_k)&=\sum_{j=2,3}V_{j1}V_{j2}
C^{\widetilde{W}}_{jk}(m_Z).
\end{align}
{We note that $\overline{C}^{\widetilde{W}}_{jk}$ appears only in the RGEs, and does not contribute to the effective operators.}

{We run down these coefficients to the hadronic scale $Q_{\text{had}}=2$~GeV using the 2-loop RGEs between the electroweak and the hadronic scales ~\cite{Nihei:1994tx} (written for a generic coefficient $C$):}

\begin{equation}
  \frac{d}{d \ln Q}C =
-\biggl[
4\frac{\alpha_s}{4\pi}+\biggl(\frac{14}{3}+\frac{4}{9}N_f
+\Delta\biggr)\frac{\alpha_s^2}{(4\pi)^2}
\biggr]C ,
\end{equation}
{where $\alpha_{s}$ and $N_{f}$ are the strong  coupling and the number of the quark flavors respectively. $\Delta$ varies from one operator to another with $\Delta=0$ for $C_{LL}$ and $\Delta=-10/3$ for $C_{RL}$. The resultant partial decay width for the $p\to K^+ \bar{\nu}_i$ mode is given by}
\begin{equation}
 \Gamma(p\to K^+\bar{\nu}_i)
=\frac{m_p}{32\pi}\biggl(1-\frac{m_K^2}{m_p^2}\biggr)^2
\vert {\cal A}(p\to K^+\bar{\nu}_i)\vert^2 ,
\end{equation}
where $m_p$ and $m_K$ are the masses of the proton and kaon,
respectively.  The
amplitude ${\cal A}(p\to K^+\bar{\nu}_i)$ is the sum of the Wilson
coefficients multiplied by the corresponding hadronic matrix elements:
\begin{align}
 {\cal A}(p\to K^+\bar{\nu}_e)&=
C_{LL}(usd\nu_e)\langle K^+\vert (us)_Ld_L\vert p\rangle
+C_{LL}(uds\nu_e)\langle K^+\vert (ud)_Ls_L\vert p\rangle ,
\nonumber \\
 {\cal A}(p\to K^+\bar{\nu}_\mu)&=  C_{RL}(usd\nu_\mu)\langle K^+\vert (us)_Rd_L\vert p\rangle
+
C_{LL}(usd\nu_\mu)\langle K^+\vert (us)_Ld_L\vert p\rangle \nonumber \\
&+C_{LL}(uds\nu_\mu)\langle K^+\vert (ud)_Ls_L\vert p\rangle ,
\nonumber \\
 {\cal A}(p\to K^+\bar{\nu}_\tau)&=
C_{RL}(usd\nu_\tau)\langle K^+\vert (us)_Rd_L\vert p\rangle
+
C_{RL}(uds\nu_\tau)\langle K^+\vert (ud)_Rs_L\vert p\rangle
\nonumber \\
&+
C_{LL}(usd\nu_\tau)\langle K^+\vert (us)_Ld_L\vert p\rangle
+C_{LL}(uds\nu_\tau)\langle K^+\vert (ud)_Ls_L\vert p\rangle .
\end{align}

{The hadronic matrix elements of the effective operators at the scale of $Q_{\text{had}}=2$~GeV have been  determined by lattice QCD computations, which we adopt  \cite{Aoki:2017puj}: }
\begin{align}
\langle K^+\vert (us)_L ^{}d_L^{}\vert p\rangle &= 0.041(2)(5)
~~\text{GeV}^2 ,
 \nonumber\\
\langle K^+\vert (ud)_L ^{}s_L^{}\vert p\rangle &= 0.139(4)(15)
~~\text{GeV}^2 ,
\nonumber \\
\langle K^+\vert (us)_R ^{}d_L^{}\vert p\rangle &= -0.049(2)(5)
~~\text{GeV}^2 ,
\nonumber \\
\langle K^+\vert (ud)_R ^{}s_L^{}\vert p\rangle &= -0.134(4)(14)
~~\text{GeV}^2 .
\end{align}

\bibliographystyle{JHEP}
\bibliography{Proton_bgu}

\providecommand{\href}[2]{#2}\begingroup\raggedright\begin{thebibliography}{10}

\bibitem{Dimopoulos:1981zb}
S.~Dimopoulos and H.~Georgi, \emph{{Softly Broken Supersymmetry and SU(5)}},
  \href{https://doi.org/10.1016/0550-3213(81)90522-8}{\emph{Nucl. Phys. B}
  {\bfseries 193} (1981) 150}.

\bibitem{Sakai:1981gr}
N.~Sakai, \emph{{Naturalness in Supersymmetric Guts}},
  \href{https://doi.org/10.1007/BF01573998}{\emph{Z. Phys. C} {\bfseries 11}
  (1981) 153}.

\bibitem{Pati:1974yy}
J.C.~Pati and A.~Salam, \emph{{Lepton Number as the Fourth Color}},
  \href{https://doi.org/10.1103/PhysRevD.10.275}{\emph{Phys. Rev. D} {\bfseries
  10} (1974) 275}.

\bibitem{Georgi:1974sy}
H.~Georgi and S.~Glashow, \emph{{Unity of All Elementary Particle Forces}},
  \href{https://doi.org/10.1103/PhysRevLett.32.438}{\emph{Phys. Rev. Lett.}
  {\bfseries 32} (1974) 438}.

\bibitem{Weinberg:1981wj}
S.~Weinberg, \emph{{Supersymmetry at Ordinary Energies. 1. Masses and
  Conservation Laws}},
  \href{https://doi.org/10.1103/PhysRevD.26.287}{\emph{Phys. Rev. D} {\bfseries
  26} (1982) 287}.

\bibitem{Sakai:1981pk}
N.~Sakai and T.~Yanagida, \emph{{Proton Decay in a Class of Supersymmetric
  Grand Unified Models}},
  \href{https://doi.org/10.1016/0550-3213(82)90457-6}{\emph{Nucl. Phys. B}
  {\bfseries 197} (1982) 533}.

\bibitem{Langacker:1980js}
P.~Langacker, \emph{{Grand Unified Theories and Proton Decay}},
  \href{https://doi.org/10.1016/0370-1573(81)90059-4}{\emph{Phys. Rept.}
  {\bfseries 72} (1981) 185}.

\bibitem{Nath:2006ut}
P.~Nath and P.~Fileviez~Perez, \emph{{Proton stability in grand unified
  theories, in strings and in branes}},
  \href{https://doi.org/10.1016/j.physrep.2007.02.010}{\emph{Phys. Rept.}
  {\bfseries 441} (2007) 191}
  [\href{https://arxiv.org/abs/hep-ph/0601023}{{\ttfamily hep-ph/0601023}}].

\bibitem{Abe:2014mwa}
{\scshape Super-Kamiokande} collaboration, \emph{{Search for proton decay via
  $p\to\nu K^+$ using 260 kiloton\textperiodcentered{}year data of
  Super-Kamiokande}},
  \href{https://doi.org/10.1103/PhysRevD.90.072005}{\emph{Phys. Rev. D}
  {\bfseries 90} (2014) 072005}
  [\href{https://arxiv.org/abs/1408.1195}{{\ttfamily 1408.1195}}].

\bibitem{Aad:2012tfa}
{\scshape ATLAS} collaboration, \emph{{Observation of a new particle in the
  search for the Standard Model Higgs boson with the ATLAS detector at the
  LHC}}, \href{https://doi.org/10.1016/j.physletb.2012.08.020}{\emph{Phys.
  Lett. B} {\bfseries 716} (2012) 1}
  [\href{https://arxiv.org/abs/1207.7214}{{\ttfamily 1207.7214}}].

\bibitem{Chatrchyan:2012ufa}
{\scshape CMS} collaboration, \emph{{Observation of a New Boson at a Mass of
  125 GeV with the CMS Experiment at the LHC}},
  \href{https://doi.org/10.1016/j.physletb.2012.08.021}{\emph{Phys. Lett. B}
  {\bfseries 716} (2012) 30} [\href{https://arxiv.org/abs/1207.7235}{{\ttfamily
  1207.7235}}].

\bibitem{Aaboud:2017dmy}
{\scshape ATLAS} collaboration, \emph{{Search for supersymmetry in final states
  with two same-sign or three leptons and jets using 36 fb$^{-1}$ of
  $\sqrt{s}=13$ TeV $pp$ collision data with the ATLAS detector}},
  \href{https://doi.org/10.1007/JHEP09(2017)084}{\emph{JHEP} {\bfseries 09}
  (2017) 084} [\href{https://arxiv.org/abs/1706.03731}{{\ttfamily
  1706.03731}}].

\bibitem{Sirunyan:2017cwe}
{\scshape CMS} collaboration, \emph{{Search for supersymmetry in multijet
  events with missing transverse momentum in proton-proton collisions at 13
  TeV}}, \href{https://doi.org/10.1103/PhysRevD.96.032003}{\emph{Phys. Rev. D}
  {\bfseries 96} (2017) 032003}
  [\href{https://arxiv.org/abs/1704.07781}{{\ttfamily 1704.07781}}].

\bibitem{Chamseddine:1982jx}
A.H.~Chamseddine, R.L.~Arnowitt and P.~Nath, \emph{{Locally Supersymmetric
  Grand Unification}},
  \href{https://doi.org/10.1103/PhysRevLett.49.970}{\emph{Phys. Rev. Lett.}
  {\bfseries 49} (1982) 970}.

\bibitem{Ellis:1979fg}
J.R.~Ellis and M.K.~Gaillard, \emph{{Fermion Masses and Higgs Representations
  in SU(5)}}, \href{https://doi.org/10.1016/0370-2693(79)90476-3}{\emph{Phys.
  Lett. B} {\bfseries 88} (1979) 315}.

\bibitem{Ellis:2002wv}
J.R.~Ellis, K.A.~Olive and Y.~Santoso, \emph{{The MSSM parameter space with
  nonuniversal Higgs masses}},
  \href{https://doi.org/10.1016/S0370-2693(02)02071-3}{\emph{Phys. Lett. B}
  {\bfseries 539} (2002) 107}
  [\href{https://arxiv.org/abs/hep-ph/0204192}{{\ttfamily hep-ph/0204192}}].

\bibitem{Babu:2014sga}
K.~Babu, I.~Gogoladze, S.~Raza and Q.~Shafi, \emph{{Flavor Symmetry Based MSSM
  (sMSSM): Theoretical Models and Phenomenological Analysis}},
  \href{https://doi.org/10.1103/PhysRevD.90.056001}{\emph{Phys. Rev. D}
  {\bfseries 90} (2014) 056001}
  [\href{https://arxiv.org/abs/1406.6078}{{\ttfamily 1406.6078}}].

\bibitem{Babu:2014lwa}
K.~Babu, I.~Gogoladze, Q.~Shafi and C.S.~\"Un, \emph{{Muon g-2, 125 GeV Higgs
  boson, and neutralino dark matter in a flavor symmetry-based MSSM}},
  \href{https://doi.org/10.1103/PhysRevD.90.116002}{\emph{Phys. Rev. D}
  {\bfseries 90} (2014) 116002}
  [\href{https://arxiv.org/abs/1406.6965}{{\ttfamily 1406.6965}}].

\bibitem{Murayama:2001ur}
H.~Murayama and A.~Pierce, \emph{{Not even decoupling can save minimal
  supersymmetric SU(5)}},
  \href{https://doi.org/10.1103/PhysRevD.65.055009}{\emph{Phys. Rev. D}
  {\bfseries 65} (2002) 055009}
  [\href{https://arxiv.org/abs/hep-ph/0108104}{{\ttfamily hep-ph/0108104}}].

\bibitem{Bajc:2002bv}
B.~Bajc, P.~Fileviez~Perez and G.~Senjanovic, \emph{{Proton decay in minimal
  supersymmetric SU(5)}},
  \href{https://doi.org/10.1103/PhysRevD.66.075005}{\emph{Phys. Rev. D}
  {\bfseries 66} (2002) 075005}
  [\href{https://arxiv.org/abs/hep-ph/0204311}{{\ttfamily hep-ph/0204311}}].

\bibitem{Bachas:1995yt}
C.~Bachas, C.~Fabre and T.~Yanagida, \emph{{Natural gauge coupling unification
  at the string scale}},
  \href{https://doi.org/10.1016/0370-2693(95)01561-2}{\emph{Phys. Lett. B}
  {\bfseries 370} (1996) 49}
  [\href{https://arxiv.org/abs/hep-th/9510094}{{\ttfamily hep-th/9510094}}].

\bibitem{Chkareuli:1998wi}
J.~Chkareuli and I.~Gogoladze, \emph{{Unification picture in minimal
  supersymmetric SU(5) model with string remnants}},
  \href{https://doi.org/10.1103/PhysRevD.58.055011}{\emph{Phys. Rev. D}
  {\bfseries 58} (1998) 055011}
  [\href{https://arxiv.org/abs/hep-ph/9803335}{{\ttfamily hep-ph/9803335}}].

\bibitem{Emmanuel-Costa:2003szk}
D.~Emmanuel-Costa and S.~Wiesenfeldt, \emph{{Proton decay in a consistent
  supersymmetric SU(5) GUT model}},
  \href{https://doi.org/10.1016/S0550-3213(03)00301-8}{\emph{Nucl. Phys. B}
  {\bfseries 661} (2003) 62}
  [\href{https://arxiv.org/abs/hep-ph/0302272}{{\ttfamily hep-ph/0302272}}].

\bibitem{Georgi:1979df}
H.~Georgi and C.~Jarlskog, \emph{{A New Lepton - Quark Mass Relation in a
  Unified Theory}},
  \href{https://doi.org/10.1016/0370-2693(79)90842-6}{\emph{Phys. Lett. B}
  {\bfseries 86} (1979) 297}.

\bibitem{Babu:2012pb}
K.~Babu, B.~Bajc and Z.~Tavartkiladze, \emph{{Realistic Fermion Masses and
  Nucleon Decay Rates in SUSY SU(5) with Vector-Like Matter}},
  \href{https://doi.org/10.1103/PhysRevD.86.075005}{\emph{Phys. Rev. D}
  {\bfseries 86} (2012) 075005}
  [\href{https://arxiv.org/abs/1207.6388}{{\ttfamily 1207.6388}}].

\bibitem{Bajc:2015ita}
B.~Bajc, S.~Lavignac and T.~Mede, \emph{{Resurrecting the minimal
  renormalizable supersymmetric SU(5) model}},
  \href{https://doi.org/10.1007/JHEP01(2016)044}{\emph{JHEP} {\bfseries 01}
  (2016) 044} [\href{https://arxiv.org/abs/1509.06680}{{\ttfamily
  1509.06680}}].

\bibitem{Ellis:2016tjc}
J.~Ellis, J.L.~Evans, A.~Mustafayev, N.~Nagata and K.A.~Olive, \emph{{The
  Super-GUT CMSSM Revisited}},
  \href{https://doi.org/10.1140/epjc/s10052-016-4437-6}{\emph{Eur. Phys. J. C}
  {\bfseries 76} (2016) 592}
  [\href{https://arxiv.org/abs/1608.05370}{{\ttfamily 1608.05370}}].

\bibitem{Ellis:2019fwf}
J.~Ellis, J.L.~Evans, N.~Nagata, K.A.~Olive and L.~Velasco-Sevilla,
  \emph{{Supersymmetric proton decay revisited}},
  \href{https://doi.org/10.1140/epjc/s10052-020-7872-3}{\emph{Eur. Phys. J. C}
  {\bfseries 80} (2020) 332}
  [\href{https://arxiv.org/abs/1912.04888}{{\ttfamily 1912.04888}}].

\bibitem{Hisano:1992mh}
J.~Hisano, H.~Murayama and T.~Yanagida, \emph{{Probing GUT scale mass spectrum
  through precision measurements on the weak scale parameters}},
  \href{https://doi.org/10.1103/PhysRevLett.69.1014}{\emph{Phys. Rev. Lett.}
  {\bfseries 69} (1992) 1014}.

\bibitem{Hisano:1992jj}
J.~Hisano, H.~Murayama and T.~Yanagida, \emph{{Nucleon decay in the minimal
  supersymmetric SU(5) grand unification}},
  \href{https://doi.org/10.1016/0550-3213(93)90636-4}{\emph{Nucl. Phys. B}
  {\bfseries 402} (1993) 46}
  [\href{https://arxiv.org/abs/hep-ph/9207279}{{\ttfamily hep-ph/9207279}}].

\bibitem{Yamada:1992kv}
Y.~Yamada, \emph{{SUSY and GUT threshold effects in SUSY SU(5) models}},
  \href{https://doi.org/10.1007/BF01650433}{\emph{Z. Phys. C} {\bfseries 60}
  (1993) 83}.

\bibitem{Hisano:1994hb}
J.~Hisano, T.~Moroi, K.~Tobe and T.~Yanagida, \emph{{Limit on the color triplet
  Higgs mass in the minimum supersymmetric SU (5) model}},
  \href{https://doi.org/10.1142/S0217732395002428}{\emph{Mod. Phys. Lett. A}
  {\bfseries 10} (1995) 2267}
  [\href{https://arxiv.org/abs/hep-ph/9411298}{{\ttfamily hep-ph/9411298}}].

\bibitem{Polonsky:1994sr}
N.~Polonsky and A.~Pomarol, \emph{{GUT effects in the soft supersymmetry
  breaking terms}},
  \href{https://doi.org/10.1103/PhysRevLett.73.2292}{\emph{Phys. Rev. Lett.}
  {\bfseries 73} (1994) 2292}
  [\href{https://arxiv.org/abs/hep-ph/9406224}{{\ttfamily hep-ph/9406224}}].

\bibitem{Lucas:1996bc}
V.~Lucas and S.~Raby, \emph{{Nucleon decay in a realistic SO(10) SUSY GUT}},
  \href{https://doi.org/10.1103/PhysRevD.55.6986}{\emph{Phys. Rev. D}
  {\bfseries 55} (1997) 6986}
  [\href{https://arxiv.org/abs/hep-ph/9610293}{{\ttfamily hep-ph/9610293}}].

\bibitem{Babu:1998wi}
K.~Babu, J.C.~Pati and F.~Wilczek, \emph{{Fermion masses, neutrino
  oscillations, and proton decay in the light of Super-Kamiokande}},
  \href{https://doi.org/10.1016/S0550-3213(99)00589-1}{\emph{Nucl. Phys. B}
  {\bfseries 566} (2000) 33}
  [\href{https://arxiv.org/abs/hep-ph/9812538}{{\ttfamily hep-ph/9812538}}].

\bibitem{Babu:2010ej}
K.~Babu, J.C.~Pati and Z.~Tavartkiladze, \emph{{Constraining Proton Lifetime in
  SO(10) with Stabilized Doublet-Triplet Splitting}},
  \href{https://doi.org/10.1007/JHEP06(2010)084}{\emph{JHEP} {\bfseries 06}
  (2010) 084} [\href{https://arxiv.org/abs/1003.2625}{{\ttfamily 1003.2625}}].

\bibitem{DiazCruz:2000mn}
J.~Diaz-Cruz, H.~Murayama and A.~Pierce, \emph{{Can supersymmetric loops
  correct the fermion mass relations in SU(5)?}},
  \href{https://doi.org/10.1103/PhysRevD.65.075011}{\emph{Phys. Rev. D}
  {\bfseries 65} (2002) 075011}
  [\href{https://arxiv.org/abs/hep-ph/0012275}{{\ttfamily hep-ph/0012275}}].

\bibitem{Enkhbat:2009jt}
T.~Enkhbat, \emph{{SU(5) unification for Yukawas through SUSY threshold
  effects}},  \href{https://arxiv.org/abs/0909.5597}{{\ttfamily 0909.5597}}.

\bibitem{Nath:1988tx}
P.~Nath and R.L.~Arnowitt, \emph{{Limits on Photino and Squark Masses From
  Proton Lifetime in Supergravity and Superstring Models}},
  \href{https://doi.org/10.1103/PhysRevD.38.1479}{\emph{Phys. Rev. D}
  {\bfseries 38} (1988) 1479}.

\bibitem{Hill:1983xh}
C.T.~Hill, \emph{{Are There Significant Gravitational Corrections to the
  Unification Scale?}},
  \href{https://doi.org/10.1016/0370-2693(84)90451-9}{\emph{Phys. Lett. B}
  {\bfseries 135} (1984) 47}.

\bibitem{Shafi:1983gz}
Q.~Shafi and C.~Wetterich, \emph{{Modification of \{GUT\} Predictions in the
  Presence of Spontaneous Compactification}},
  \href{https://doi.org/10.1103/PhysRevLett.52.875}{\emph{Phys. Rev. Lett.}
  {\bfseries 52} (1984) 875}.

\bibitem{Hall:1992kq}
L.J.~Hall and U.~Sarid, \emph{{Gravitational smearing of minimal supersymmetric
  unification predictions}},
  \href{https://doi.org/10.1103/PhysRevLett.70.2673}{\emph{Phys. Rev. Lett.}
  {\bfseries 70} (1993) 2673}
  [\href{https://arxiv.org/abs/hep-ph/9210240}{{\ttfamily hep-ph/9210240}}].

\bibitem{Dasgupta:1995js}
T.~Dasgupta, P.~Mamales and P.~Nath, \emph{{Effects of gravitational smearing
  on predictions of supergravity grand unification}},
  \href{https://doi.org/10.1103/PhysRevD.52.5366}{\emph{Phys. Rev. D}
  {\bfseries 52} (1995) 5366}
  [\href{https://arxiv.org/abs/hep-ph/9501325}{{\ttfamily hep-ph/9501325}}].

\bibitem{Paige:2003mg}
F.E.~Paige, S.D.~Protopopescu, H.~Baer and X.~Tata, \emph{{ISAJET 7.69: A Monte
  Carlo event generator for pp, anti-p p, and e+e- reactions}},
  \href{https://arxiv.org/abs/hep-ph/0312045}{{\ttfamily hep-ph/0312045}}.

\bibitem{Baer:2002fv}
H.~Baer, C.~Balazs and A.~Belyaev, \emph{{Neutralino relic density in minimal
  supergravity with coannihilations}},
  \href{https://doi.org/10.1088/1126-6708/2002/03/042}{\emph{JHEP} {\bfseries
  03} (2002) 042} [\href{https://arxiv.org/abs/hep-ph/0202076}{{\ttfamily
  hep-ph/0202076}}].

\bibitem{Group:2009ad}
{\scshape CDF, D0} collaboration, \emph{{Combination of CDF and D0 Results on
  the Mass of the Top Quark}},
  \href{https://arxiv.org/abs/0903.2503}{{\ttfamily 0903.2503}}.

\bibitem{Ajaib:2013zha}
M.~Adeel~Ajaib, I.~Gogoladze, Q.~Shafi and C.S.~Un, \emph{{A Predictive Yukawa
  Unified SO(10) Model: Higgs and Sparticle Masses}},
  \href{https://doi.org/10.1007/JHEP07(2013)139}{\emph{JHEP} {\bfseries 07}
  (2013) 139} [\href{https://arxiv.org/abs/1303.6964}{{\ttfamily 1303.6964}}].

\bibitem{Pierce:1996zz}
D.M.~Pierce, J.A.~Bagger, K.T.~Matchev and R.-j.~Zhang, \emph{{Precision
  corrections in the minimal supersymmetric standard model}},
  \href{https://doi.org/10.1016/S0550-3213(96)00683-9}{\emph{Nucl. Phys. B}
  {\bfseries 491} (1997) 3}
  [\href{https://arxiv.org/abs/hep-ph/9606211}{{\ttfamily hep-ph/9606211}}].

\bibitem{Nakamura:2010zzi}
{\scshape Particle Data Group} collaboration, \emph{{Review of particle
  physics}}, \href{https://doi.org/10.1088/0954-3899/37/7A/075021}{\emph{J.
  Phys. G} {\bfseries 37} (2010) 075021}.

\bibitem{Aaboud:2017vwy}
{\scshape ATLAS} collaboration, \emph{{Search for squarks and gluinos in final
  states with jets and missing transverse momentum using 36 fb$^{-1}$ of
  $\sqrt{s}=13$ TeV pp collision data with the ATLAS detector}},
  \href{https://doi.org/10.1103/PhysRevD.97.112001}{\emph{Phys. Rev. D}
  {\bfseries 97} (2018) 112001}
  [\href{https://arxiv.org/abs/1712.02332}{{\ttfamily 1712.02332}}].

\bibitem{Aaij:2012nna}
{\scshape LHCb} collaboration, \emph{{First Evidence for the Decay $B_s^0 \to
  \mu^+ \mu^-$}},
  \href{https://doi.org/10.1103/PhysRevLett.110.021801}{\emph{Phys. Rev. Lett.}
  {\bfseries 110} (2013) 021801}
  [\href{https://arxiv.org/abs/1211.2674}{{\ttfamily 1211.2674}}].

\bibitem{Amhis:2012bh}
{\scshape Heavy Flavor Averaging Group} collaboration, \emph{{Averages of
  B-Hadron, C-Hadron, and tau-lepton properties as of early 2012}},
  \href{https://arxiv.org/abs/1207.1158}{{\ttfamily 1207.1158}}.

\bibitem{Asner:2010qj}
{\scshape Heavy Flavor Averaging Group} collaboration, \emph{{Averages of
  $b$-hadron, $c$-hadron, and $\tau$-lepton properties}},
  \href{https://arxiv.org/abs/1010.1589}{{\ttfamily 1010.1589}}.

\bibitem{Hinshaw:2012aka}
{\scshape WMAP} collaboration, \emph{{Nine-Year Wilkinson Microwave Anisotropy
  Probe (WMAP) Observations: Cosmological Parameter Results}},
  \href{https://doi.org/10.1088/0067-0049/208/2/19}{\emph{Astrophys. J. Suppl.}
  {\bfseries 208} (2013) 19} [\href{https://arxiv.org/abs/1212.5226}{{\ttfamily
  1212.5226}}].

\bibitem{Akrami:2018vks}
{\scshape Planck} collaboration, \emph{{Planck 2018 results. I. Overview and
  the cosmological legacy of Planck}},
  \href{https://doi.org/10.1051/0004-6361/201833880}{\emph{Astron. Astrophys.}
  {\bfseries 641} (2020) A1}
  [\href{https://arxiv.org/abs/1807.06205}{{\ttfamily 1807.06205}}].

\bibitem{Baer:2012up}
H.~Baer, V.~Barger, P.~Huang, A.~Mustafayev and X.~Tata, \emph{{Radiative
  natural SUSY with a 125 GeV Higgs boson}},
  \href{https://doi.org/10.1103/PhysRevLett.109.161802}{\emph{Phys. Rev. Lett.}
  {\bfseries 109} (2012) 161802}
  [\href{https://arxiv.org/abs/1207.3343}{{\ttfamily 1207.3343}}].

\bibitem{Ellis:2015rya}
J.~Ellis, J.L.~Evans, F.~Luo, N.~Nagata, K.A.~Olive and P.~Sandick,
  \emph{{Beyond the CMSSM without an Accelerator: Proton Decay and Direct Dark
  Matter Detection}},
  \href{https://doi.org/10.1140/epjc/s10052-015-3842-6}{\emph{Eur. Phys. J. C}
  {\bfseries 76} (2016) 8} [\href{https://arxiv.org/abs/1509.08838}{{\ttfamily
  1509.08838}}].

\bibitem{Brink:2005ej}
{\scshape CDMS-II} collaboration, \emph{{Beyond the CDMS-II dark matter search:
  SuperCDMS}}, {\emph{eConf} {\bfseries C041213} (2004) 2529}
  [\href{https://arxiv.org/abs/astro-ph/0503583}{{\ttfamily
  astro-ph/0503583}}].

\bibitem{Aramaki:2016spe}
{\scshape SuperCDMS} collaboration, \emph{{Recent results from the second
  CDMSlite run and overview of the SuperCDMS SNOLAB project}},
  \href{https://doi.org/10.22323/1.268.0030}{\emph{PoS} {\bfseries DSU2015}
  (2016) 030}.

\bibitem{Akerib:2018lyp}
{\scshape LUX-ZEPLIN} collaboration, \emph{{Projected WIMP sensitivity of the
  LUX-ZEPLIN dark matter experiment}},
  \href{https://doi.org/10.1103/PhysRevD.101.052002}{\emph{Phys. Rev. D}
  {\bfseries 101} (2020) 052002}
  [\href{https://arxiv.org/abs/1802.06039}{{\ttfamily 1802.06039}}].

\bibitem{Aprile:2020vtw}
{\scshape XENON} collaboration, \emph{{Projected WIMP sensitivity of the
  XENONnT dark matter experiment}},
  \href{https://doi.org/10.1088/1475-7516/2020/11/031}{\emph{JCAP} {\bfseries
  2011} (2020) 031} [\href{https://arxiv.org/abs/2007.08796}{{\ttfamily
  2007.08796}}].

\bibitem{Akerib:2016lao}
{\scshape LUX} collaboration, \emph{{Results on the Spin-Dependent Scattering
  of Weakly Interacting Massive Particles on Nucleons from the Run 3 Data of
  the LUX Experiment}},
  \href{https://doi.org/10.1103/PhysRevLett.116.161302}{\emph{Phys. Rev. Lett.}
  {\bfseries 116} (2016) 161302}
  [\href{https://arxiv.org/abs/1602.03489}{{\ttfamily 1602.03489}}].

\bibitem{Tanaka:2011uf}
{\scshape Super-Kamiokande} collaboration, \emph{{An Indirect Search for WIMPs
  in the Sun using 3109.6 days of upward-going muons in Super-Kamiokande}},
  \href{https://doi.org/10.1088/0004-637X/742/2/78}{\emph{Astrophys. J.}
  {\bfseries 742} (2011) 78} [\href{https://arxiv.org/abs/1108.3384}{{\ttfamily
  1108.3384}}].

\bibitem{IceCube:2009iyf}
{\scshape IceCube} collaboration, \emph{{Limits on a muon flux from neutralino
  annihilations in the Sun with the IceCube 22-string detector}},
  \href{https://doi.org/10.1103/PhysRevLett.102.201302}{\emph{Phys. Rev. Lett.}
  {\bfseries 102} (2009) 201302}
  [\href{https://arxiv.org/abs/0902.2460}{{\ttfamily 0902.2460}}].

\bibitem{Khachatryan:2014rra}
{\scshape CMS} collaboration, \emph{{Search for dark matter, extra dimensions,
  and unparticles in monojet events in proton\textendash{}proton collisions at
  $\sqrt{s} = 8$ TeV}},
  \href{https://doi.org/10.1140/epjc/s10052-015-3451-4}{\emph{Eur. Phys. J. C}
  {\bfseries 75} (2015) 235} [\href{https://arxiv.org/abs/1408.3583}{{\ttfamily
  1408.3583}}].

\bibitem{Gogoladze:2011be}
I.~Gogoladze, S.~Raza and Q.~Shafi, \emph{{Light stop from b\textendash{}$\tau$
  Yukawa unification}},
  \href{https://doi.org/10.1016/j.physletb.2011.11.026}{\emph{Phys. Lett. B}
  {\bfseries 706} (2012) 345}
  [\href{https://arxiv.org/abs/1104.3566}{{\ttfamily 1104.3566}}].

\bibitem{Babu:2018tfi}
K.~Babu, B.~Bajc and S.~Saad, \emph{{Resurrecting Minimal Yukawa Sector of SUSY
  SO(10)}}, \href{https://doi.org/10.1007/JHEP10(2018)135}{\emph{JHEP}
  {\bfseries 10} (2018) 135}
  [\href{https://arxiv.org/abs/1805.10631}{{\ttfamily 1805.10631}}].

\bibitem{Ellis:1983qm}
J.R.~Ellis, J.~Hagelin, D.V.~Nanopoulos and K.~Tamvakis, \emph{{Observable
  Gravitationally Induced Baryon Decay}},
  \href{https://doi.org/10.1016/0370-2693(83)91557-5}{\emph{Phys. Lett. B}
  {\bfseries 124} (1983) 484}.

\bibitem{Belyaev:1982ik}
V.~Belyaev and M.~Vysotsky, \emph{{More About Proton Decay Due to d = 5
  Operators}}, \href{https://doi.org/10.1016/0370-2693(83)90879-1}{\emph{Phys.
  Lett. B} {\bfseries 127} (1983) 215}.

\bibitem{Alonso:2014zka}
R.~Alonso, H.-M.~Chang, E.E.~Jenkins, A.V.~Manohar and B.~Shotwell,
  \emph{{Renormalization group evolution of dimension-six baryon number
  violating operators}},
  \href{https://doi.org/10.1016/j.physletb.2014.05.065}{\emph{Phys. Lett. B}
  {\bfseries 734} (2014) 302}
  [\href{https://arxiv.org/abs/1405.0486}{{\ttfamily 1405.0486}}].

\bibitem{Nihei:1994tx}
T.~Nihei and J.~Arafune, \emph{{The Two loop long range effect on the proton
  decay effective Lagrangian}},
  \href{https://doi.org/10.1143/PTP.93.665}{\emph{Prog. Theor. Phys.}
  {\bfseries 93} (1995) 665}
  [\href{https://arxiv.org/abs/hep-ph/9412325}{{\ttfamily hep-ph/9412325}}].

\bibitem{Aoki:2017puj}
Y.~Aoki, T.~Izubuchi, E.~Shintani and A.~Soni, \emph{{Improved lattice
  computation of proton decay matrix elements}},
  \href{https://doi.org/10.1103/PhysRevD.96.014506}{\emph{Phys. Rev. D}
  {\bfseries 96} (2017) 014506}
  [\href{https://arxiv.org/abs/1705.01338}{{\ttfamily 1705.01338}}].

\end{thebibliography}\endgroup

\end{document}